%% file: main.tex
\documentclass[aps, pra, reprint, nofootinbib]{revtex4-2}
\usepackage[colorlinks,linkcolor=red,anchorcolor=blue,citecolor=blue]{hyperref}
\usepackage{graphicx} 
\usepackage{amsmath}
\usepackage{amssymb}
\usepackage{bbm}
\usepackage{bm}
\usepackage{multirow}
\usepackage{mathtools}
\usepackage{subcaption}
\usepackage[dvipsnames]{xcolor}
\usepackage[font=small,labelfont=bf,justification=raggedright]{caption}
\usepackage{tikz}

\usepackage{pgfplots}
\pgfplotsset{compat=1.17}
\pgfplotsset{scaled y ticks=false}

\usetikzlibrary{quantikz2}
\newtheorem{theorem}{Theorem}[section]

\newtheorem{lemma}[theorem]{Lemma}

\begin{document}
\title{Adaptive Circuit Learning of Born Machine: Towards Realization of Amplitude Embedding and Quantum Data Loading}
\author{Chun-Tse Li$^{1,5}$}
\author{Hao-Chung Cheng$^{1,2,3,4,5}$}
\email{haochung.ch@gmail.com}
\affiliation{$^1$Department of Electrical Engineering and Graduate Institute of \\ Communication Engineering, National Taiwan University, Taipei 106, Taiwan (R.O.C.)}
\affiliation{$^2$Department of Mathematics, National Taiwan University, Taipei 106, Taiwan (R.O.C.)}
\affiliation{$^3$Center for Quantum Science and Engineering,  National Taiwan University}
\affiliation{$^4$Physics/Mathematics Division, National Center for Theoretical Sciences, Taipei 10617, Taiwan (R.O.C.)}
\affiliation{$^5$Hon Hai Quantum Computing Research Center, Taipei, Taiwan}


\begin{abstract}
    Quantum data loading plays a central role in quantum algorithms and quantum information processing. Many quantum algorithms hinge on the ability to prepare arbitrary superposition states as a subroutine, with claims of exponential speedups often predicated on access to an efficient data-loading oracle. In practice, constructing a circuit to prepare a generic $n$-qubit quantum state typically demands computational efforts scaling as $\mathcal{O}(2^n)$, posing a significant challenge for quantum algorithms to outperform their classical counterparts. To address this critical issue, various hybrid quantum-classical approaches have been proposed. However, many of these solutions favor simplistic circuit architectures, which are susceptible to substantial optimization challenges.

    In this study, we harness quantum circuits as Born machines to generate probability distributions. Drawing inspiration from methods used to investigate electronic structures in quantum chemistry and condensed matter physics, we propose a framework called Adaptive Circuit Learning of Born Machine, which dynamically expands the ansatz circuit. Our algorithm is designed to selectively integrate two-qubit entangled gates that best capture the intricate entanglement present within the target state. 
    Empirical experiments underscore the efficacy of our approach in encoding real-world data through amplitude embedding, demonstrating not only compliance with but also enhancement over the performance benchmarks set by prior research.
\end{abstract}

\maketitle

\section{Introduction}
Over the past two decades, intensive research within the quantum computing community has spurred rapid advancements in quantum algorithms, yielding promising applications across diverse fields of study. These include solving systems of linear equations~\cite{harrow2009quantum, childs2009equation, clader2013preconditioned, lloyd2013quantum}, Hamiltonian simulation~\cite{childs2012hamiltonian, berry2015simulating, berry2015hamiltonian, low2017optimal, low2019hamiltonian, gilyen2019quantum}, the quantum approximation optimization algorithm ~\cite{farhi2014quantum, zhou2020quantum, hadfield2019quantum}, quantum walks~\cite{childs2003exponential, magniez2007quantum}, quantum principal component analysis~\cite{lloyd2014quantum}, and quantum support vector machines~\cite{rebentrost2014quantum}. However, many of the claimed exponential speedups over their classical counterparts are contingent upon the assumption that  designate quantum states
can be efficiently prepared~\cite{aaronson2015read}. This task, often referred to as ``quantum data loading," plays a pivotal role in a wide variety of quantum algorithms~\cite{harrow2009quantum,childs2009equation,childs2012hamiltonian,low2017optimal,low2019hamiltonian,gilyen2019quantum,orus2019quantum,lloyd2013quantum, lloyd2014quantum,rebentrost2014quantum}, yet it incurs significant overhead. In reality, studies have shown that preparing generic quantum states requires either $\mathcal{O}(2^n/n)$ circuit depth with $n$ qubits~\cite{barenco1995elementary, vartiainen2004efficient, shende2005synthesis, plesch2011quantum} or $\Theta(n)$ circuit depth with $\mathcal{O}(2^n)$ ancillary qubits~\cite{sun2023asymptotically, rosenthal2021query, zhang2022quantum}. Only certain $d$-sparse quantum states,  i.e., quantum states with only $d$ non-zero amplitudes, can be prepared in $\Theta(\log(nd))$ circuit depth. However, this still necessitates $\mathcal{O}(nd\log(d))$ ancillary qubits~\cite{zhang2022quantum}, which might exceed the capabilities of current NISQ computers.

Despite the formidable complexity of general state preparation, it remains an open question whether certain highly structured and low-entanglement quantum states can be approximately constructed with $\mathcal{O}(\textnormal{poly}(n))$ overhead. Recent studies suggest that the quantum state representations of real-valued, smooth, and differentiable functions exhibit slow entanglement growth as the number of qubits increases~\cite{holmes2020efficient,garcia2021quantum}, potentially allowing for a reasonable increase in overhead as precision improves. Moreover, some approaches utilizing Matrix Product States (MPS)~\cite{holmes2020efficient,garcia2021quantum,iaconis2024quantum,dilip2022data} have been put forward to encode these smooth and  differentiable functions into quantum states, demonstrating that MPS can efficiently represent these distributions with high precision and relatively low bond dimensions. These findings provide evidence that highly structured states might exhibit low complexity and can be constructed using efficient quantum circuits, thereby avoiding the exponential overhead typically associated with general state preparation.

Previously, several hybrid quantum-classical algorithms have been proposed to address quantum data loading.
Quantum Generative Adversarial Network (QGAN)~\cite{dallaire2018quantum, lloyd2018quantum, zoufal2019quantum} employs adversarial training to minimize the distance between the distribution produced by a quantum generator and the target distribution. Quantum Circuit Born Machine (QCBM)~\cite{liu2018differentiable} utilizes the squared maximum mean discrepancy (MMD) loss to measure the distance between the model and target distribution in the kernel feature space, and they construct the entanglement layer based on the Chow--Liu tree~\cite{chow1968approximating}. Data Driven Quantum Circuit Learning (DDQCL)~\cite{benedetti2019generative} applies negative log likelihood as a loss function to train the quantum generator. 
In Ref.~\cite{dilip2022data}, an image data is encoded into a quantum state utilizing the MPS-based variational anasatz. Additionally, in Ref.~\cite{selig2023deepqprep}, instead of relying on variational quantum circuits like other approaches, reinforcement learning techniques have been adopted to search for the optimal gates to reconstruct the quantum state from a universal quantum gate set.

While promising experimental results have been demonstrated on some simple distributions, it is unknown whether these approaches can be generalized to more complex and higher dimensional distribution. In fact, numerical studies reveal their inability to deliver satisfactory outcomes on more intricate probability distributions.
The noticeable degradation in performance with an increasing problem size can likely be attributed to the simplistic circuit structures originally designed for preparing the ground states of small molecules \cite{kandala2017hardware}. 
Although these structures are straightforward to implement and agnostic to the problem at hand, they suffer from an exponential decrease in gradient magnitude as the number of qubits and circuit depth increase~\cite{mcclean2018barren}. 
Furthermore, neglecting to incorporate the underlying structure and properties of the distribution into the circuit design raises concerns about whether the resources are optimally utilized to approximately prepare the target states.

Conversely, the adaptive circuit structure ansatz, and in particular, the ADAPT-VQE method, as outlined in~\cite{grimsley2019adaptive, tang2021qubit}, has attracted considerable interest from the quantum computing community. The ADAPT-VQE iteratively builds the ansatz by selectively incorporating the unitary operators from a predetermined operator pool based on their gradient magnitudes. This innovative strategy results in the creation of remarkably compact ansatz, delivering superior precision when compared to previous methodologies referenced in~\cite{taube2006new, romero2018strategies, lee2018generalized}. Additionally, numerical evidence indicates that ansatz produced by this algorithm are resistant to barren plateaus, offering a more favorable scenario for optimization tasks. Recently, there are some similar algorithms that have been proposed to tackle other challenges, such as quantum machine learning~\cite{bilkis2021semi} and quantum circuit compilation~\cite{cincio2018learning, cincio2021machine}. However, it is unclear whether adaptive circuit learning can be effectively applied to quantum data loading.

In this study, we adopt the concept of the adaptive circuit learning framework and introduce the framework ``Adaptive Circuit Learning of Born Machine'' (ACLBM). This framework is designed to iteratively build an ansatz circuit chosen based on the target distribution.  
Furthermore, we propose a \textit{new} operator pool containing the $SO(4)$ local unitary operators and numerically demonstrate its effectiveness for various types of data distributions. 
We extensively benchmark the performance of our model against a variety of data distributions, as well as compare it with existing approaches in the literature. Specifically, we demonstrate ACLBM's capability to represent real-world data, such as images, through amplitude embedding, encoding image information into the $2^n$ amplitudes of an $n$-qubit state. Our numerical experiments reveal that our method surpasses current techniques~\cite{zoufal2019quantum, liu2018differentiable, benedetti2019generative,dilip2022data} by requiring fewer parameters and yielding more accurate results, showcasing an exceptional capability to discern the intrinsic correlations within image data with a parameter count of only $\mathcal{O}(\textnormal{poly}(n))$.
Moreover, we propose some potential applications of our framework on quantum algorithms, including constructing the data loading oracle in Hamiltonian simulation and Monte Carlo pricing in quantum finance.

\section{Framework}
In this section, we lay the groundwork for understanding the quantum data loading problem and the Adaptive Circuit Learning of Born Machines (ACLBM) algorithm. We begin by introducing basic notation and formalizing the quantum data loading problem in Sec.~\ref{subsec:quantum_data_loading_problem}. In Sec.~\ref{subsec:data_representation}, we discuss how to encode different types of data into a quantum state, including discrete and continuous probability distributions, as well as classical data. Additionally, we explain how to discretize continuous probability distributions for encoding into a quantum state. In Sec.~\ref{subsec:loss_criterion}, we outline the loss criterion used in ACLBM, which guides the evaluation of the algorithm’s performance. Finally, in Sec.~\ref{subsec:ACLBM_algorithm}, we provide a comprehensive overview of the ACLBM algorithm, breaking down its core components: the operator pool, circuit initialization, gradient calculation, operator selection, and optimization.

\subsection{Quantum Data Loading Problem}
\label{subsec:quantum_data_loading_problem}
In the context of quantum data loading, our goal is to approximately prepare a state corresponding to a given probability distribution $\{p(x)\}_{x=0}^{2^n-1}$:

\begin{equation}
\label{eq:target_state}
|\Psi^\star\rangle=\sum_{x=0}^{2^n-1}e^{i\phi_x}\sqrt{p(x)} \ |x\rangle,
\end{equation}
such that, in accordance with the Born rule, the probability of measuring a specific computational basis $\{|x\rangle\}_{x=0}^{2^n-1}$ is

\begin{equation}
\label{eq:prob}
|\langle x|\Psi^\star\rangle|^2=p(x).
\end{equation}

The objective is to identify a unitary transformation $U(\bm{\theta})$ that effectively reproduces the target probability distribution upon measuring in the computational basis,

\begin{equation}
\text{Tr}\left(|x\rangle\langle x|U(\bm{\theta})|0\rangle\langle 0|^{\otimes n}U^\dagger(\bm{\theta})\right)=q_{\bm{\theta}}(x)\approx p(x).
\end{equation}

Note that we allow for an arbitrary phase factor $e^{i\phi_x}$ in the functional form of $|\Psi^\star\rangle$. This flexibility is permissible since, in most quantum algorithm subroutines, it suffices to satisfy the form in Eq.~(\ref{eq:target_state}) rather than strictly adhering to $\sum_{x=0}^{2^n-1}\sqrt{p(x)}|x\rangle$. This allowance enables us to search within a larger functional space and reduces the need for full quantum state tomography, as only projective measurements in the computational basis are required to determine the probability distribution in Eq.~(\ref{eq:prob}).

\subsection{Data Representation}
\label{subsec:data_representation}

\begin{figure}[t]
\centering
\begin{tikzpicture}
    \begin{axis}[
        title={Discretization of continuous probability (log normal)},
        xlabel={$x$},
        ylabel={Probability},
        legend pos=north east,
        ymajorgrids=true,
        grid style=dashed,
        ymin=0, ymax=0.08, 
        ytick={0.02,0.04,0.06}, 
        yticklabels={0.02,0.04,0.06}, 
    ]
    
    \addplot[
        domain=1:64,
        samples=100,
        smooth,
        thick,
        black
        ]
        {1/(x*0.3*sqrt(2*pi))*exp(-(ln(x)-3)^2/(2*0.3^2))};
    
    \addplot[
        ybar,
        bar width=0.9, 
        fill=teal,
        fill opacity=0.5,
        draw=gray, 
        line width=0.5pt, 
        ]
        table [x index=0, y index=1, col sep=space] {data/pmf_data.txt};
    
    \end{axis}
\end{tikzpicture}    
\caption{Discretization of a log-normal probability distribution. The black curve represents the probability density function of a log-normal distribution over the interval $[1, 64]$ with parameters $\mu=3$ and $\sigma=0.3$. The blue-green bars illustrate the discretized probability mass function obtained by applying the formula in Eq.~(\ref{eq:discretized_PMF}).}
\end{figure}

We now discuss how to transform the representation of classical data into the form in Eq.(\ref{eq:target_state}). For a generic continuous probability distribution, possibly on an infinite support, discretization and truncation are necessary to encode the distribution into a quantum state. Given a probability density function $f(x)$, the discretized (and unnormalized) probability mass function $\tilde{p}(x)$ could be obtained by evaluating $f(x)$ at integer points. For some probability density functions $f(x)$, where the tail probability decays exponentially, it may be reasonable to truncate the distribution to a finite interval $[a, b)$. We assume $b-a$ is an integer of the form $2^n$ without loss of generality, as zero-padding can be used to satisfy this condition. The probability mass function $p(x)$ for Eq.~\eqref{eq:target_state} could be constructed by normalizing $\tilde{p}(x)$ and applying the appropriate shift $x\mapsto x+a$:
\begin{equation}
\label{eq:discretized_PMF}
    p(x)=\frac{\tilde{p}(x+a)}{\sum_{x=0}^{2^n-1} \tilde{p}(x+a)}.
\end{equation}
To increase the precision of discretization, we could introduce an $m$-bit auxiliary variable $s\in[0,1)$ such that the interval of each discretized point can be further subdivided into $2^m$ subintervals:
\begin{equation}
\label{eq:discretized_cont_prob}
    p(x,s)=p(x+s)=\frac{\tilde{p}(x+s+a)}{\sum_{x=0}^{2^n-1}\sum_{y=0}^{2^m-1}\tilde{p}(x+\frac{y}{2^m}+a)},
\end{equation} 
Eq.~(\ref{eq:target_state}) can then be slightly modified by introducing an auxiliary register that represents the floating point $s$:
\begin{equation}
    |\Psi^\star\rangle=\sum_{x=0}^{2^n-1}\sum_{y=0}^{2^m-1} e^{i\phi_{x,y}}\sqrt{p(x,y)}|x\rangle|y\rangle,
\end{equation}
where we implicitly apply the change of variable $s=y/2^m$. By introducing the auxiliary qubits, the discretization error to calculate the expectation value $\varepsilon=\vert\int_a^bg(x)f(x)dx-\sum_{xy}g(x+a+\frac{y}{2^m})p(x,y)\vert\sim\mathcal{O}(2^{-m})$ decays exponentially, allowing us to approximate the continuous distribution arbitrarily well by introducing more and more auxiliary bits $m$.

In cases where the underlying distribution is unknown \textit{a priori}, it may be empirically reconstructed from $N$ independent and identical samples $\mathcal{D}=\{x_1,...,x_N\}$ as follows:
\begin{equation}
        p(x)=\frac{1}{N}\sum_{i=1}^N \mathbbm{1}\{x=x_i\},  \\[0.2cm]
\end{equation}
for a discrete underlying distribution, and
\begin{equation}
    p(x,s)=\frac{1}{N}\sum_{i=1}^N \mathbbm{1}\{x+s-2^{-m-1} \leq x_i \leq x+s+2^{-m-1}\}
\end{equation}
for a continuous distribution with discretization precision up to $2^{-m}$. Note that, given a sufficiently large sample size $N$, the empirical probability distribution converges to the underlying distribution by the Law of Large Numbers.

For classical data, we assume it can be represented by a $2^n$-dimensional vector that can be normalized with a unit $\ell_2$ norm. For example, a $256\times256$ image could be flattened into a $65536$ dimensional vector,  which can then be normalized to form a probability distribution. This technique, used to encode classical data into a quantum state as shown in Eq.~\eqref{eq:target_state}, is known as ``Amplitude Embedding" and is commonly adopted in the quantum machine learning literature.

\subsection{Loss criterion}
\label{subsec:loss_criterion}

To accurately quantify the discrepancy between approximated and target quantum states, it is crucial to select an appropriate distance measure. While fidelity, $F(\rho,\sigma)=(\text{Tr}\sqrt{\sqrt{\rho}\sigma\sqrt{\rho}})^2$ and trace distance, $T(\rho,\sigma)=\frac{1}{2}\Vert\rho-\sigma\Vert_1$, are commonly employed for this purpose, their computational complexity can make them impractical for large-scale applications. Furthermore, in scenarios where the relative phase $e^{i\phi_x}$ is not of primary concern — meaning the focus is solely on the absolute amplitude $\sqrt{p(x)}$ — these measures may not always be necessary. This is because many quantum states can satisfy a given probability distribution in Eq.~(\ref{eq:prob}), whereas these measures assume convergence to a specific state.

Having access only to the approximated probability distribution $q_{\bm{\theta}}(x)$, the Kullback--Leibler (KL) divergence, and the Fisher--Rao metric emerge as suitable metrics to quantify the discrepancy between two distributions:
\begin{align}
&\mathcal{L}_1(\bm{\theta})=D_{\text{KL}}(p\Vert q_{\bm{\theta}})=\sum_{x}p(x)\log \frac{p(x)}{q_{\bm{\theta}}(x)}.
\\
&\mathcal{L}_2(\bm{\theta})=\arccos\langle\sqrt{p},\sqrt{q_{\bm{\theta}}}\rangle=\arccos\sum_{x}\sqrt{p(x)q_{\bm{\theta}}(x)}
\label{eq:fisher-rao}
\end{align}
Numerically, the KL divergence effectively characterizes the distance between distributions across any dimensionality without suffering from diminishing values as the problem size increases.
On the other hand, Fisher--Rao metric can effectively capture the distance better when the distributions are highly unstructured, e.g., the flattened image data.

\begin{figure*}[t]
    \centering
        \includegraphics[width=\textwidth]{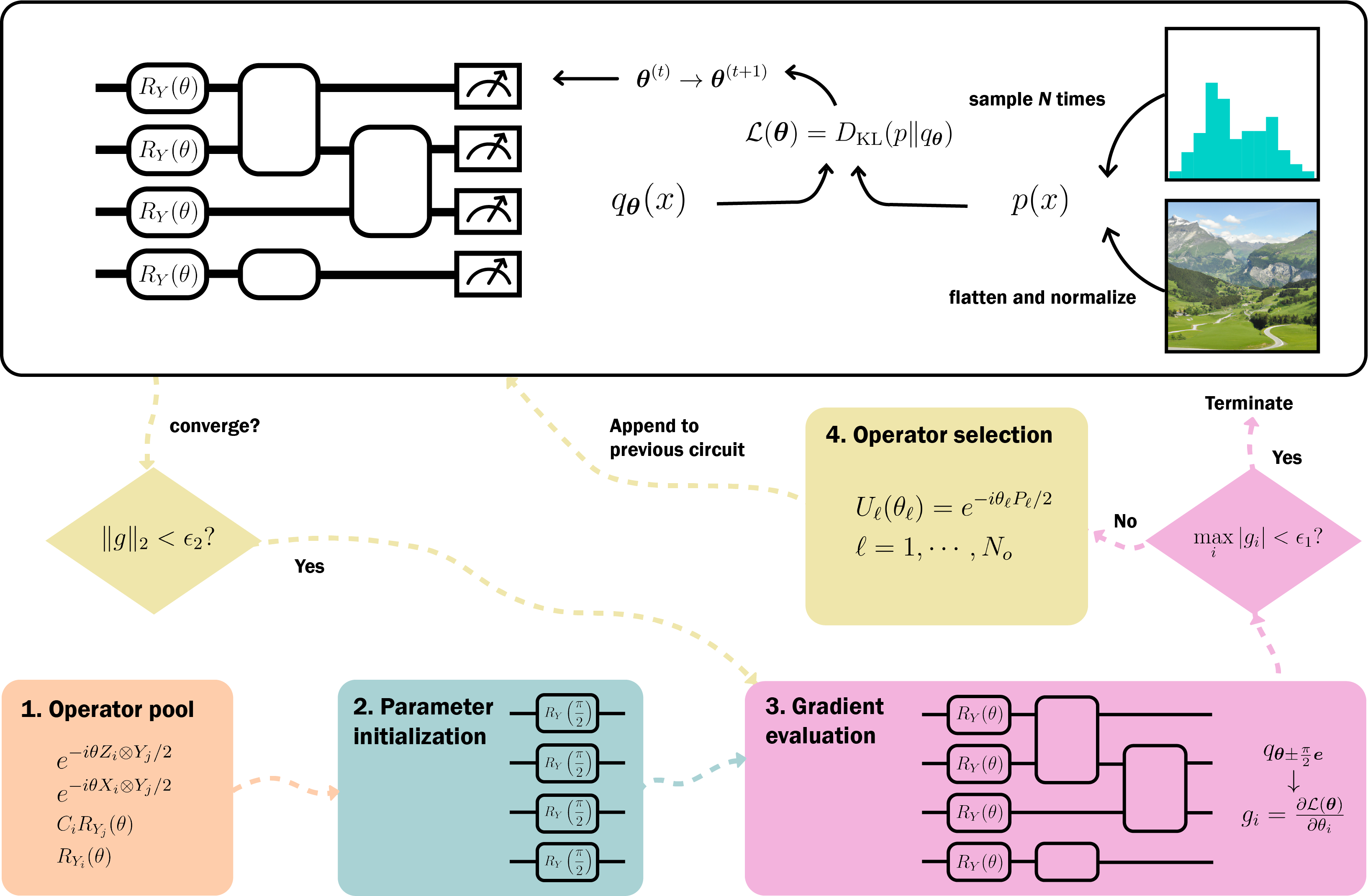}
        \caption{Graphical illustration of the Adaptive Circuit Learning of Born Machine (ACLBM) framework. The upper portion of the diagram outlines the quantum data loading framework, where a quantum circuit, functioning as a Born machine, generates a probability distribution. This distribution is then utilized to compute the Kullback-Leibler (KL) divergence, serving as the loss function, which is subsequently minimized by a classical optimizer to update the circuit parameters. The lower portion delineates the adaptive learning workflow. At each iteration, a set of $N_o$ operators is selected and incorporated into the current circuit configuration. A reoptimization of all parameters follows. The optimization cycle concludes once the convergence criterion \textbf{--} specifically, when the $L_2$ norm of the gradient vector $\Vert\bm{g}\Vert_2$ \textbf{--} is less than a prescribed threshold $\epsilon_2$. The entire procedure terminates when the maximum gradient among the selected operators is below $\epsilon_1$, ensuring the algorithm's progression towards a precise solution without perturbing the state, as gradients are evaluated at $\theta_i=0 $.}
        \label{fig:flow_chart}
\end{figure*}

\subsection{Adaptive circuit learning algorithm}
\label{subsec:ACLBM_algorithm}

In this section, we describe the steps of the adaptive circuit learning algorithm in detail. 
Readers are referred to Fig.~\ref{fig:flow_chart} for a schematic workflow.
\\[0.25cm]
\textbf{Operator pool.} The first step in the adaptive circuit learning algorithm is to establish a self-defined operator pool. Grimsley et al.’s foundational ADAPT-VQE work~\cite{grimsley2019adaptive} chooses a pool comprising fermionic single and double excitation operators, drawing from the UCC ansatz~\cite{romero2018strategies}. This method capitalizes on unitary coupled cluster theory, integrating physical intuition into the operator selection process. However, our problem’s context lacks such physical motivations for constructing the operator pool, prompting a shift towards alternative methodologies. The qubit-ADAPT-VQE, proposed by Tang et al.~\cite{tang2021qubit}, presents a more fitting approach for our scenario. Tang et al.~\cite{tang2021qubit} have shown that a complete pool, comprising a minimal number of $2n-2$ Pauli operators, is capable of generating any arbitrary quantum state through the application of unitaries derived from these Pauli operators. 

Despite this, our numerical simulations indicate that the specific pool of $2n-2$ Pauli operators recommended in~\cite{tang2021qubit} does not consistently produce satisfactory results.  We surmise that this might be attributed to the fact that nearest-neighbor unitary operations, such as $e^{i\theta Z_i\otimes Y_{i+1}}$ ($i$ is the index of qubit), struggle to effectively propagate interactions between distant qubits as the problem size increases. This issue becomes even more apparent when the distribution is sparse. 

To address this limitation, we propose a {new} operator pool that includes $SO(4)$ local unitary operators acting on all pairs of qubits. 
As indicated in previous literature~\cite{tucci2005introduction,zhang2003geometric}, these operators form a basis of $\mathfrak{so}(4)$, the Lie algebra of the special orthogonal group $SO(4)$:
\begin{equation}
    \frac{i}{2}Q^\dagger\{X_1,Y_1,Z_1,X_2,Y_2,Z_2\}Q
\end{equation}
where
\begin{equation}
    Q=\frac{1}{\sqrt{2}}\begin{pmatrix}
        1 & 0 & 0 & i \\ 0 & i & 1 & 0 \\ 0 & i & -1 & 0 \\ 1 & 0 & 0 & -i
    \end{pmatrix}
\end{equation}
A straightforward calculation shows that this set of operators is actually
\begin{equation}
    -\frac{i}{2}\{Z_1Y_2, Y_1, X_1Y_2, Y_2, -Y_1Z_2, Y_1X_2\}  
\end{equation}
which consists of six one-body and two-body Pauli operators with real matrix entries. By exponentiating these basis operators for all pairs of qubits, we generate six types of rotations in $SO(4)$. Additionally, we include controlled-RY gates to further enhance the expressivity of the operator pool $\mathcal{P}$:

\begin{align}
    e^{-i\theta Z_i\otimes Y_j / 2} &=
    \begin{pmatrix}
    \cos\frac{\theta}{2} & -\sin\frac{\theta}{2} & 0 & 0 \\ \sin\frac{\theta}{2} & \cos\frac{\theta}{2} & 0 & 0 \\ 0 & 0 & \cos\frac{\theta}{2} & \sin\frac{\theta}{2}  \\ 0 & 0 & -\sin\frac{\theta}{2} & \cos\frac{\theta}{2}
    \end{pmatrix}\hspace{0.3cm} \forall i\neq j, 
    \nonumber
\end{align}
\begin{align}
    e^{-i\theta X_i \otimes Y_j / 2} &= 
    \begin{pmatrix}
    \cos\frac{\theta}{2} & 0 & 0 & -\sin\frac{\theta}{2} \\ 0 & \cos\frac{\theta}{2} & \sin\frac{\theta}{2} & 0 \\ 0 & -\sin\frac{\theta}{2} & \cos\frac{\theta}{2} & 0 \\ \sin\frac{\theta}{2} & 0 & 0 & \cos\frac{\theta}{2}
    \end{pmatrix}\hspace{0.3cm} \forall i\neq j,
    \nonumber
\end{align}
\begin{align}
    C_iR_{Y_j}(\theta) &=
    \begin{pmatrix}
    1 & 0 & 0 & 0 \\ 0 & 1 & 0 & 0 \\ 0 & 0 & \cos\frac{\theta}{2} & -\sin\frac{\theta}{2} \\ 0 & 0 & \sin\frac{\theta}{2} & \cos\frac{\theta}{2}
    \end{pmatrix}\hspace{0.3cm} \forall i\neq j,
    \nonumber
\end{align}
\begin{align}
    R_{Y_i}(\theta) &=
    \begin{pmatrix}
    \cos\frac{\theta}{2} & -\sin\frac{\theta}{2} \\ \sin\frac{\theta}{2} & \cos\frac{\theta}{2}
    \end{pmatrix} \hspace{0.8cm}\forall i.
\end{align}
The indices $i, j$ label the qubits on which the Pauli operators act.

The main reason to restrict our attention to real unitary operators is primarily motivated by the observation that, in cases where the generator is real (consider $Z_iX_j$ as an example), constraining the state to the real domain leads to a vanishing gradient. For a detailed proof and further discussion on this phenomenon, please see Appendix~\ref{Appendix:Zero gradient operators}.
\\[0.25cm]
\textbf{Ansatz initialization.} In the preceding discussion, 
the KL divergence and Fisher--Rao metric were selected as the loss functions, with their gradients incorporating the approximated probability $q_{\bm{\theta}}(x)$ in its denominator. 
This can potentially result in an explosion of values for both the loss function and its gradient, particularly when the approximated distribution $q_{\bm{\theta}}(x)$ has zero values inside the support of the target distribution $p(x)$. To circumvent this issue, we initialize our quantum state to be an equal superposition, achieved by applying a set of $R_Y(\theta)|_{\theta=\frac{\pi}{2}}$ gates to each qubit. Furthermore, we allow these initial parameters to be trainable in the subsequent optimization phase.
\\[0.25cm]

\begin{figure*}[ht!]
    \centering
    \begin{subfigure}{0.4\textwidth}
        \centering
        \includegraphics[width=\textwidth]{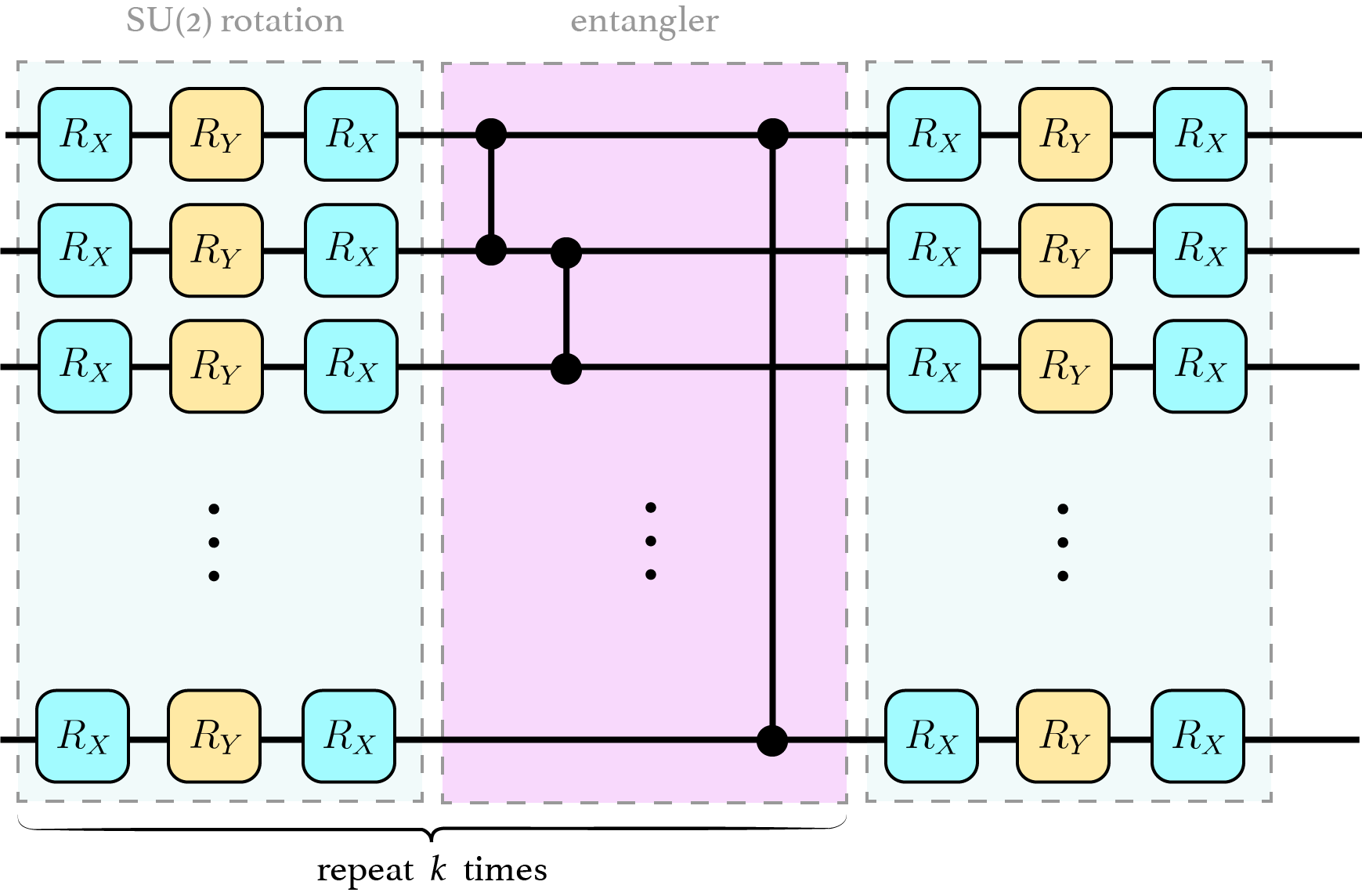}
        \caption{circuit structure 1}
        \label{fig:circuit1}
    \end{subfigure}
    \hspace*{1cm}
    \begin{subfigure}{0.216\textwidth}
        \centering
        \includegraphics[width=\textwidth]{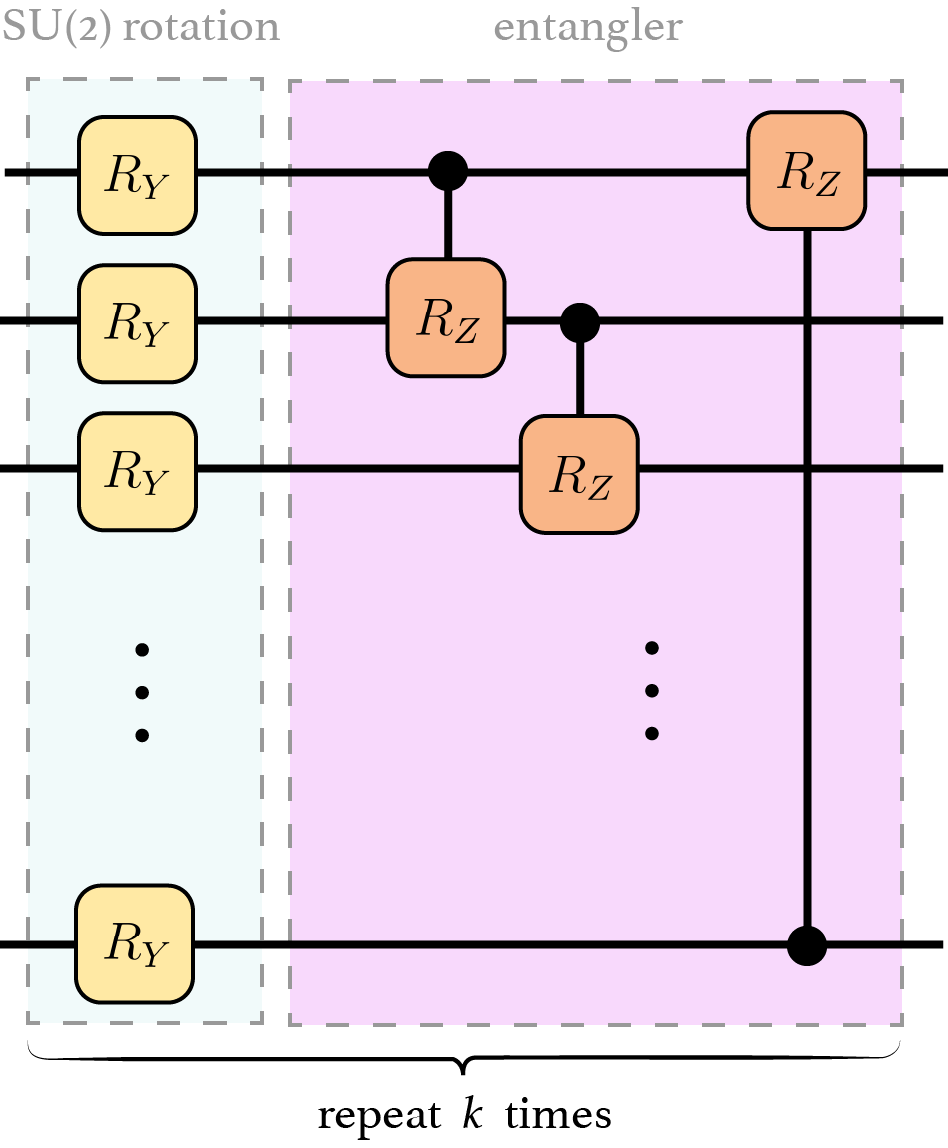}
        \caption{circuit structure 2}
        \label{fig:circuit2}
    \end{subfigure}
    \hspace*{1cm}
    \begin{subfigure}{0.175\textwidth}
        \centering
        \includegraphics[width=\textwidth]{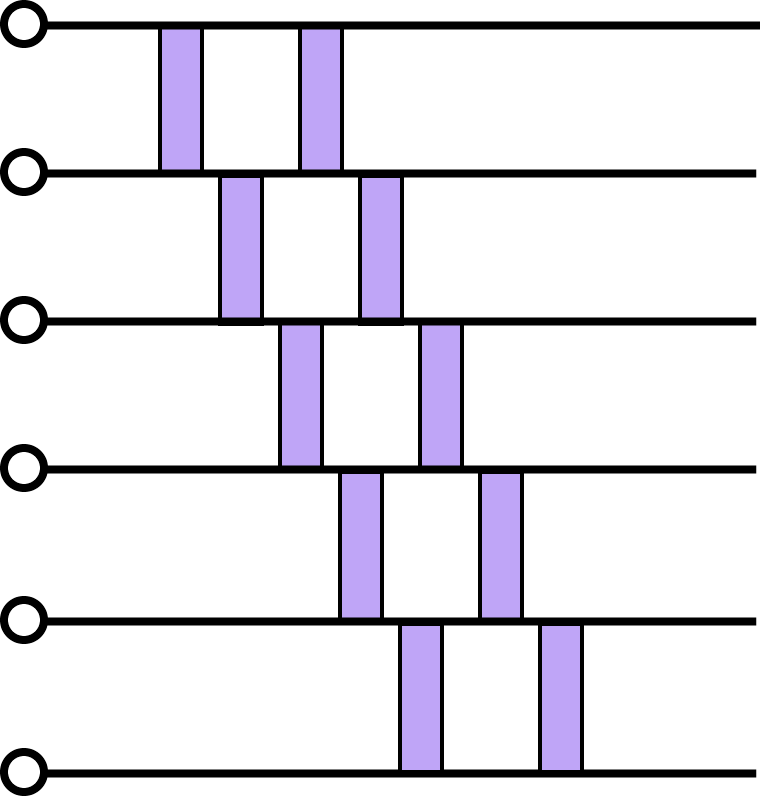}
        \vspace*{0.4cm}
        \caption{MPS circuit}
        \label{fig:mps_circuit}
    \end{subfigure}
    \caption{
    Different circuit structures for QGAN \cite{zoufal2019quantum}, QCBM \cite{liu2018differentiable}, DDQCL \cite{benedetti2019generative} and MPS \cite{dilip2022data}.
    }
\end{figure*}

\noindent
\textbf{Gradient evaluation.}  
In the second step, we compute the gradient of the operators within the pool. The gradients are evaluated at the points where the appended unitary operator is the identity, i.e., with parameters set to zero, to avoid perturbing the preceding state. Previous studies~\cite{romero2018strategies,tang2021qubit} show that the gradient evaluated at identity ($\theta=0$) can be derived by measuring the expectation value of the commutator between the Hamiltonian $H$ and the generator $P$ of unitary operator $e^{-i\theta P/2}$, i.e.,
\begin{equation}
    \frac{\partial\langle  H \rangle_{\bm{\theta}}}{\partial \theta}\biggr|_{\theta=0}=-\frac{i}{2}\langle\psi(\bm{\theta})|[H,P]|\psi(\bm{\theta})\rangle.
\end{equation}
where we collect the model parameters as a vector $\bm{\theta}$, and $\theta$ denotes the parameter corresponding to the evaluated unitary operator $e^{-i\theta P/2}$.

However, in our case, we do not possess a defined Hamiltonian. As detailed in Appendix~\ref{Appendix:Gradient evaluation}, we establish that the gradient can be analytically calculated using the parameter-shift rule~\cite{mitarai2018quantum, schuld2019evaluating}. Therefore, instead of measuring the commutator, we can append the unitary $U(\theta)\in\mathcal{P}$ (with $\theta=0$) directly to the preceding circuit and utilize three quantum evaluations to compute the exact gradient: \\
For the KL divergence:
\begin{equation}
    \frac{\partial \mathcal{L}(\bm{\theta})}{\partial \theta}\Bigg\vert_{\theta=0}=-\sum_x \frac{p(x)}{q_{\bm{\theta}}(x)}\biggl(q_{\bm{\theta}+\frac{\pi}{2}\bm{e}}(x)-q_{\bm{\theta}-\frac{\pi}{2}\bm{e}}(x)\biggr).
\end{equation}
And for Fisher--Rao metric:
\begin{equation}
    \frac{\partial \mathcal{L}(\bm{\theta})}{\partial \theta}\Bigg\vert_{\theta=0}=\frac{\sum_x\sqrt{\frac{p(x)}{q_{\bm{\theta}}(x)}}\biggl(q_{\bm{\theta}+\frac{\pi}{2}\bm{e}}(x)-q_{\bm{\theta}-\frac{\pi}{2}\bm{e}}(x)\biggr)}{2\sqrt{1-\langle \sqrt{p},\sqrt{q_{\bm{\theta}}}\rangle^2}},
\end{equation}
where $\bm{e}$ denotes the unit vector with nonzero entry at the location of $\theta$, indicating a shift in the parameter $\theta$ by $\pm\frac{\pi}{2}$.

Notice that if the largest gradient is less than the threshold, say $\epsilon_1$, we simply terminate our algorithm.
\\[0.25cm]

\noindent\textbf{Operator selection.} 
Upon calculating the gradients for each operator, we proceed to select the top $N_o$ operators based on the largest gradient magnitudes and then append them to the previous circuit in either randomized order or descending order of their gradient magnitudes. While previous literature, such as the works by Romero et al.~\cite{romero2018strategies} and Tang et al.~\cite{tang2021qubit}, has typically set $N_o$ to 1, our numerical experiments indicate that a moderately larger $N_o$ often enhances performance. This approach not only yields better results but also reduces the necessity of repeatedly evaluating the gradients within the operator pool. Additionally, in scenarios involving extremely sparse distributions, a larger $N_o$ becomes crucial to ensure convergence of the algorithm.
\\[0.25cm]
\textbf{Optimization.} We incorporate the chosen operator into our ansatz and proceed to optimize all the parameters until convergence is achieved. The convergence criterion is defined as the gradient norm 
$\|\bm{g}\|_2$ falling below a specified threshold $\epsilon_2$.

\vspace{0.2cm}

Note that in the subsequent sections, we will refer to a parameter update as an ``epoch", and an operator selection as an ``iteration". To clarify,  a single iteration includes both the operator selection and the parameter optimization phases. Parameter optimization is conducted after the selected operators are appended to the circuit. Additionally, a single iteration may consist of multiple epochs (parameter updates).

\begin{figure*}[ht!]
    \centering
        \includegraphics[width=\textwidth]{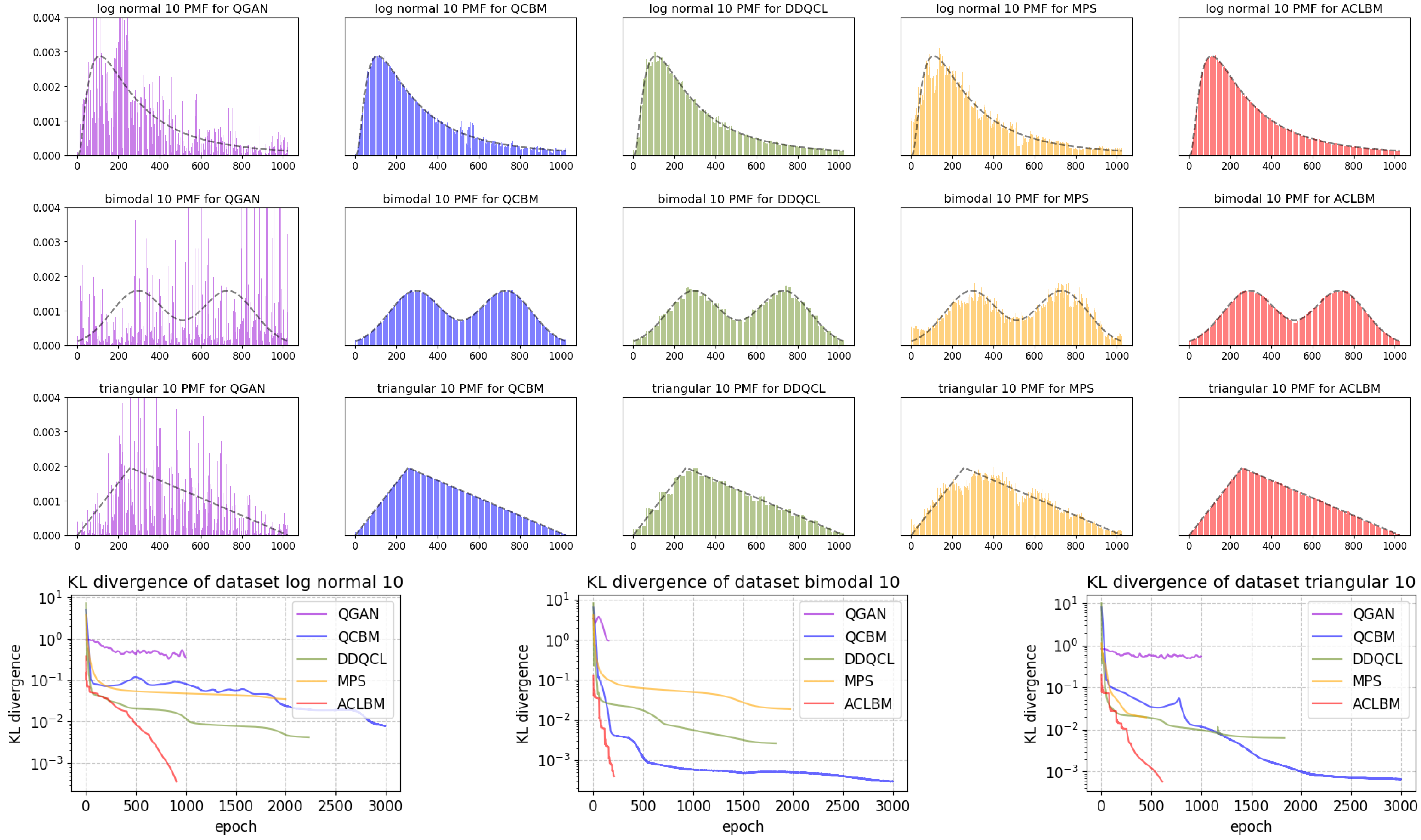}
        \caption{Benchmark comparisons for the generic probability distribution across different models. This figure illustrates the benchmarking of various quantum generative models, with QGAN \cite{zoufal2019quantum} (purple), QCBM \cite{liu2018differentiable} (blue), DDQCL \cite{benedetti2019generative} (green), MPS (yellow) \cite{dilip2022data}, and the proposed ACLBM (red). The top row represents the results for the log-normal distribution, followed by the bimodal distribution in the second row, and the triangular distribution in the third. The bottom row depicts the KL divergence across the number of epochs. We employ the circuit structure 2 in Fig.~\ref{fig:circuit2} on QCBM and DDQCL with  totaling $100$ parameters. While for QGAN, we use the circuit structure 1 in Fig.~\ref{fig:circuit1}. In contrast, the proposed ACLBM model demonstrates superior accuracy as well as parameter efficiency with only 99, 33, and 75 parameters for the respective distributions.} 
        \label{fig:generic dist}
\end{figure*}

\section{Numerical experiments}

In this section, we present an extensive study based on numerical simulations conducted across various datasets. To ensure a fair comparison, we closely followed the experimental settings outlined in the original works~\cite{zoufal2019quantum, liu2018differentiable, benedetti2019generative, dilip2022data}, while also fine-tuning each model and circuit architecture to enhance performance on different types of datasets.

For GAN \cite{zoufal2019quantum}, QCBM \cite{liu2018differentiable}, and DDQCL \cite{benedetti2019generative}, we considered two types of circuit architectures. The first architecture consists of a sequence of $k+1$ layers of single-qubit rotations $R_X(\theta_{i,1})R_Y(\theta_{i,2})R_X(\theta_{i,3})$ alternated with $k$ layers of entanglement layer. The second architecture comprises $k$ layers of single-qubit roataions $R_Y(\theta_i)$, alternated with $k$ layers of two-qubit controlled-rotations $CR_Z(\theta_{i})$.
For QCBM \cite{liu2018differentiable}, the entanglement layers are structured based on the Chow--Liu Tree~\cite{chow1968approximating}. 
In contrast, the entanglement layer for QGAN \cite{zoufal2019quantum} and DDQCL \cite{benedetti2019generative} involves a simpler arrangement where controlled-$Z$ and controlled-$R_Z$ gates connect each qubit $i$ with its neighbor $(i+1) \text{ mod }n$. Graphical illustration of these two circuit architectures could be found in Fig.~\ref{fig:circuit1} and Fig.~\ref{fig:circuit2}. 

For a MPS-like circuit, the structure is shown in Fig.~\ref{fig:mps_circuit}. Each purple block represents a two-qubit gate, and in Ref.~\cite{dilip2022data}, it is proposed to parameterize the two-qubit block using $\exp\big(-i\sum_{i,j}\theta_{i,j}\sigma_i\otimes\sigma_j\big)$. However, this block is, in fact, difficult and impractical to implement on quantum circuit since the two-qubit Pauli operators are not all pairwise commute with each other. Instead, we adopt the Trotterized version of the original proposal $\prod_{i,j}\exp(-i\theta_{i,j}\sigma_i\otimes\sigma_j)$, to avoid the difficulties associated with implementation.

In order to rule out the detrimental effect on the finite shot noise, which could obscure the true performance of different models, all experiments are conducted under the assumption that the exact probability distribution of the model, $q_{\bm{\theta}}(x)$, is accessible. This is equivalent to performing an infinite number of measurements.
For each experiment,
we fine-tune the model hyperparameters, circuit architectures, and loss functions to maximize the capacity of each model for different data distributions and ensure a fair comparison. We present and plot the best results in the following discussion. In each trial, the comparative models use Gaussian initialization with $\mu=0$ and $\sigma=\pi/8$, while our model adopts the parameter initialization strategy outlined in the previous section. Note that the figures in the following section illustrate the relationship between KL divergence and the number of epochs. For additional clarity, figures showing KL divergence vs. the number of measurements are provided in Appendix~\ref{Appendix:experiment_results} for reference.

All implementations were carried out using PennyLane~\cite{bergholm2018pennylane}, interfaced with PyTorch~\cite{paszke2019pytorch} and we utilize ADAM optimizer~\cite{kingma2014adam} to optimize the model parameters. 

\subsection{Generic probability distributions}

We assess of our algorithm on a range of generic probability distributions
---log normal, bimodal and triangular distribution, 
\begin{align}
    &f(x)=\frac{1}{x\sqrt{2\pi\sigma^2}}e^{-\frac{(\ln x-\mu)^2}{2\sigma^2}},
    \\[0.2cm]
    &f(x)=\frac{1}{\sqrt{8\pi\sigma_1^2}}e^{-\frac{(x-\mu_1)^2}{2\sigma_1^2}}+\frac{1}{\sqrt{8\pi\sigma_2^2}}e^{-\frac{(x-\mu_2)^2}{2\sigma_2^2}}
    \\[0.2cm]
    &f(x)=
    \left\{\begin{array}{ll}
        \frac{2(x-l)}{(u-l)(m-l)}, & l\leq x\leq m \\[0.2cm]
        \frac{2(u-x)}{(u-l)(u-m)}, & m\leq x\leq u
    \end{array}\right.
\end{align}
which were initially utilized as benchmarks in works by Benedetti et al.~\cite{benedetti2019generative} and Liu et al.~\cite{liu2018differentiable}. We adopt the discretization method discussed in Section.~\ref{subsec:data_representation}. It is noteworthy that these distributions are highly structured and can be uniquely specified by a small number of parameters; consequently, it remains unclear whether there are efficient techniques specifically tailored for preparing these distributions. 

Our analysis focuses on a 10-qubit dataset, examining a log-normal distribution with parameters $\mu=5.5 ,\sigma=0.9$, a bimodal distribution with $\sigma_1=\sigma_2=128, \mu_1=\frac{2}{7}\cdot 1024, \mu_2=\frac{5}{7}\cdot 1024$, and a triangular distribution with a lower limit $l=0$, upper limit $u=1023$ and mode $m=256$. For QCBM and DDQCL, we adopt circuit structure 2 from Fig.~\ref{fig:circuit2}, with circuit depth $k=5$ and a total of 100 parameters. For QGAN, we use circuit structure 1 from Fig.~\ref{fig:circuit1}, with circuit depth $k=10$ and a totoal of 330 parameters. For the MPS circuit, we employ 2 MPS layers with a total of 270 parameters.

Moreover, the numerical results indicate that QCBM suffers from an exponential decay of the MMD loss on dense data distributions as the problem size increases, leading to the vanishing gradient problem. To address this, we adopt the logarithmic version of the MMD loss in QCBM, which significantly improves performance compared to the original MMD loss. For DDQCL and MPS, we use the standard KL divergence as the loss criterion, while QGAN employs the adversarial loss function presented in Ref.~\cite{zoufal2019quantum}. As for our model, the number of operators chosen in each epoch is set to $N_o=3$, and the termination criterion $\epsilon_1=0.001$ and $\epsilon_2=0.005$.

Our numerical simulations indicate that each model can accurately prepare quantum states for datasets with a smaller number of qubits, as detailed in Appendix~\ref{Appendix:experiment_results}. Specifically, for the 3-qubit datasets, the QGAN model could achieve a KL divergence of around $10^{-2}$ to $10^{-3}$, while QCBM, DDQCL, MPS and ACBLM achieve approximately $10^{-3}$ to $10^{-6}$, or even lower. However, as the scale of the problem increases, the performance of the benchmark models start deteriorating, only reaching a KL divergence of around $10^{-2}$ to $10^{-3}$. In contrast, our model, ACLBM, demonstrates robustness, consistently achieving an excellent match with the target distribution, with a KL divergence of about $10^{-4}$ across all datasets. A detailed comparison of 10-qubit datasets is presented in Fig.~\ref{fig:generic dist}.

We also want to highlight that the states discussed in this subsection are typically smooth or piecewise smooth and require fewer resources to generate compared to the distributions of real images in Section~\ref{subsec:real_images}. Consequently, we conjecture that these types of piecewise smooth distributions might be more easily managed by ACLBM.

\subsection{Bars and Stripes}

\begin{figure}[ht!]
    \includegraphics[width=0.45\textwidth]{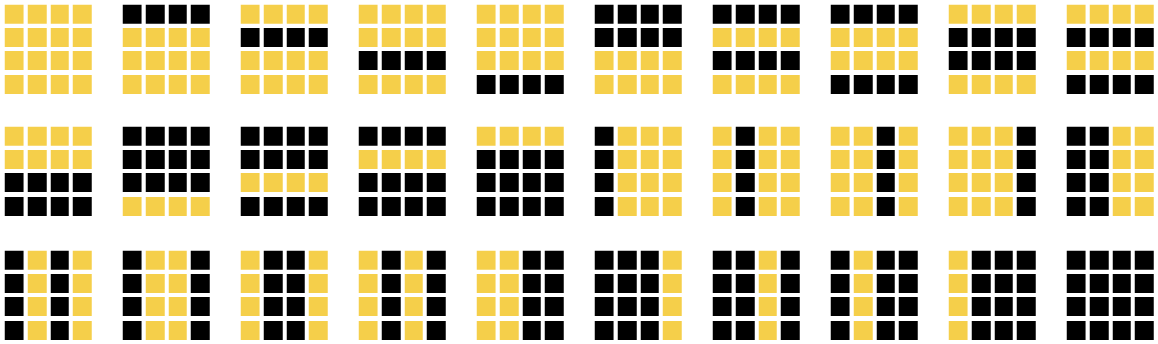}
    \caption{Representative patterns from the $4\times 4$ Bars and Stripes (BAS) dataset. This subset includes all permissible configurations of horizontal bars and vertical stripes. Among the $2^{16}=65536$ possible patterns for a $4\times 4$ grid, only 30 constitute valid BAS patterns.} 
    \label{fig:bas illustration}
\end{figure}

\begin{figure*}[ht!]
    \centering
        \includegraphics[width=\textwidth]{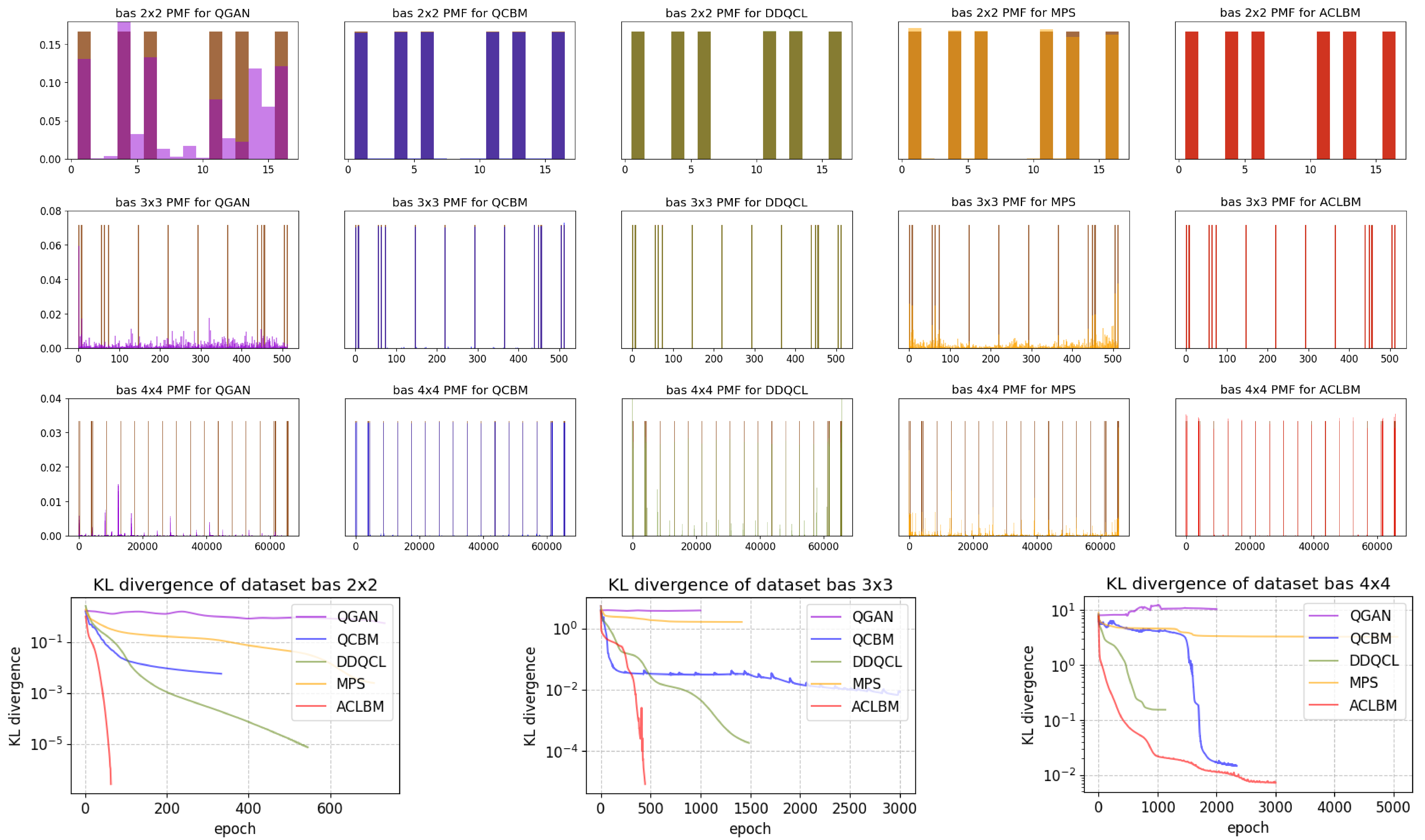}
        \caption{Benchmark comparisons on the Bars and Stripes (BAS) dataset. From top to bottom rows, we present BAS $2\times 2$, BAS $3\times 3$, BAS $4\times 4$, and the KL divergence respectively. In the histogram, the colors purple, blue, green, yellow and red represent the distributions generated by QGAN, QCBM, DDQCL, MPS and ACLBM, respectively, while the brown color denotes the target distribution.
        For the $2\times 2$ BAS dataset, almost all models can perfectly fit the distribution. However, in the $3\times 3$ BAS dataset, QGAN's performance begins to deteriorate as adversarial training struggles to capture the features of the 9-qubit data. MPS also fails to improve as the problem scales up, suggesting that the MPS ansatz may not be suitable for learning sparse distributions.  In contrast, our model (ACLBM) maintains robustness and exhibits a rapid convergence rate, utilizing $10, 80$ and $400$ parameters on the respective distributions, even for data with 65,536 dimensions.} 
        \label{fig:bas_comparison}
    \end{figure*}

The Bars and Stripes (BAS) dataset \cite{mackay2003information, han2018unsupervised, benedetti2019generative} is a set of binary images that are foundational for benchmarking generative models in both classical deep learning and quantum machine learning. Images in BAS display unique patterns of horizontal bars or vertical stripes within a grid, with `on' pixels (1's) forming the bars or stripes and `off' pixels (0's) as the background. Due to its binary and combinatorial nature, the BAS dataset serves as a prime candidate for evaluating the ability of generative algorithms to learn and replicate distinct, structured patterns, thus providing insight into the learning model's capabilities to capture data distributions effectively. All patterns of BAS $4\times4$ dataset is shown in Fig.~\ref{fig:bas illustration}.

The BAS dataset serves as an excellent testing ground for evaluating the entangling capabilities of quantum circuits. This is because finding a close approximation of BAS distributions is nearly impossible using only product states. Our model reflects this challenge; in each iteration, a higher number of operators must be selected to ensure satisfactory convergence.

In our investigation, we examine the $2\times 2$, $3\times 3$, and $4\times 4$ configurations of the BAS dataset, which correspond to quantum data loading problems involving 4, 9, and 16 qubits, respectively. For the QGAN, QCBM and DDQCL, we adopt the circuit structure 1 from Fig.~\ref{fig:circuit1}, with circuit depths of $k=4$ layers for the $2\times 2$ case, $k=10$ layers for the $3\times 3$ case, and $k=20$ layers for the $4\times 4$ case. The total number of adjustable parameters for these configurations are 60, 297, and 1008, respectively. For the MPS model, the circuit depth is set to $k=2$, $k=3$, $k=3$ for the three datasets, corresponding to 90, 360, and 675 parameters, respectively. In contrast, ACLBM employs a variable number of operators, selecting $N_o=10$ for the $2\times 2$ case, $N_o=80$ for the $3\times 3$ case, and $N_o=400$ for the $4\times 4$ case in each optimization iteration. It is worth noting that QCBM does not exhibit exponential decay in gradient values for sparse distributions. Therefore, we use the original MMD loss as presented in Ref.~\cite{liu2018differentiable}.

Our numerical experiments, shown in Fig.~\ref{fig:bas_comparison}, indicate that QCBM, DDQCL, and ACLBM are all capable of finding robust approximations of the BAS distribution, demonstrating consistency across different experimental trials. In contrast, QGAN and MPS face challenges when approximating the distribution at larger problem sizes. MPS shows limited improvement as the problem size increases, while QGAN often experiences an unstable training process, frequently converging to random quantum states rather than the desired distribution. This highlights the inherent difficulties and instabilities in adversarial learning, which often require careful tuning of hyperparameters and strategic initialization to ensure robustness across varying trials.

\subsection{Real images}
\label{subsec:real_images}

\begin{figure*}[ht!]
    \centering
    \includegraphics[width=\textwidth]{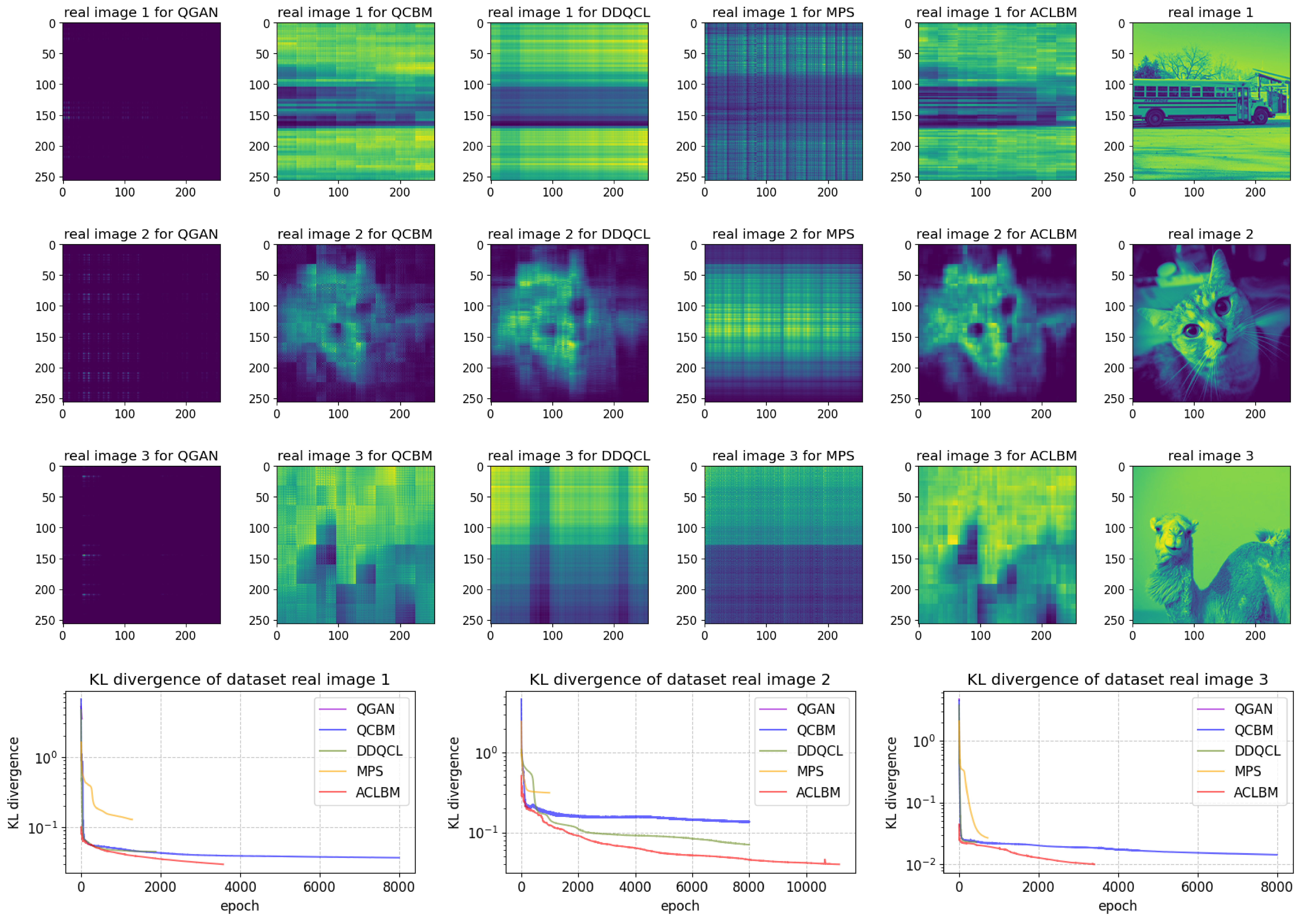}
    \caption{Benchmark comparisons on the real image dataset. For QGAN, QCBM, and DDQCL, we employ circuit structure 2 from~\ref{fig:circuit2} with circuit depths $k=20, 30$ and $20$, respectively. The results demonstrate that only QCBM, DDQCL, and ACLBM produce visually recognizable features, while QGAN, MPS, and the other models fail to do so. QGAN suffers from catastrophic optimization issues, with unstable generator and discriminator losses that cause the KL divergence to increase as iterations progress. MPS also fails to capture the underlying image distribution, resulting in unrecognizable outputs.}
    \label{fig:real_images}
\end{figure*}

We demonstrate ACLBM's capability for approximating amplitude embedding, a crucial task in quantum machine learning, where image datasets are encoded into the $2^n$ amplitudes of quantum states. In contrast to previously considered distributions which are either sparse or could be specified by few number of parameters, image datasets intrinsically present a greater challenge for quantum representation due to their encoding of exponentially many information.

We evaluate three images, each of $256 \times 256$ pixels, corresponding to a 16-qubit system. These images are transformed into 65536-dimensional vectors and normalized to form a probability distribution $\{p(x)\}_{x=0}^{65535}$. For QGAN, QCBM and DDQCL, we employ the circuit structure 2 from Fig.~\ref{fig:circuit2} with circuit depths $k=20$, $k=30$ and $k=20$ for real images 1, 2, and 3 \footnote{We chose different circuit depths based on observed performance: increasing the depth beyond 20 layers for real images 1 and 3 did not improve performance and, in some cases, led to a noticeable decline. However, for real image 2, increasing the depth from 20 to 30 layers significantly enhanced the model’s performance. Circuit depth selection is problem-dependent, so we recommend testing various depths to achieve optimal results.}, respectively, resulting in a total of 640, 960, and 640 parameters. For MPS circuit, a circuit depth $k=3$ is employed with a total 675 parameters. In our methodology, $N_o=3$ generators are selected in each iteration to construct the circuit. Notably, for this data distribution, QCBM also encounters an exponential decay in gradient values, so we use the logarithmic version of the MMD loss to mitigate this problem and significantly improve model performance. Additionally, we find that the Fisher--Rao metric in Eq.~\eqref{eq:fisher-rao} more effectively captures the intricate image data distribution than the KL divergence. Consequently, we adopt the Fisher--Rao metric as the loss criterion for ACLBM.

Numerical experiments shown in Fig.~\ref{fig:real_images} demonstrate that only QCBM, DDQCL, and our model (ACLBM) can capture distributions recognizable to the human eye. Other approaches, such as QGAN and MPS, fail to approximate the distribution of real-image datasets, resulting in outputs that lack distinguishable features. Although the images generated by our model are slightly blurred, they retain visually identifiable features, demonstrating a clear approximation of the original distribution. Notably, among all the models capable of producing human-recognizable results, our model best captures the intricate image distributions. These findings highlight the importance of well-designed circuit architectures for accurately approximating complex distributions in quantum machine learning.

It is worth noting that the Maximum Mean Discrepancy (MMD) loss function used by QCBM is less effective as a loss criterion for dense data distributions. This may be attributed to the mathematical form of the MMD loss, which tends to decay exponentially when the amplitude at each point $x$ is roughly the same. The MMD loss is defined as:
\begin{align}
    \text{MMD} &= \bigg\Vert \sum_xp(x)\phi(x)-\sum_{x}q_{\bm{\theta}}(x)\phi(x)\bigg\Vert 
    \nonumber \\
    &=\mathbbm{E}_{x,y\sim p}[K(x,y)]-2\mathbbm{E}_{x\sim p, y\sim q_{\bm{\theta}}}[K(x,y)] 
    \nonumber \\[0.2cm]
    &\hspace{0.25cm}+ \mathbbm{E}_{x,y \sim q_{\bm{\theta}}}[K(x,y)]
\end{align}
where 
\begin{equation}
    K(x,y)=\phi(x)^T\phi(y)=\exp\left(-\frac{|x-y|}{2\sigma^2}\right)
\end{equation}
is the radial basis function (RBF) kernel peaked around $x=y$. Since RBF is a exponential function of $|x-y|$, only exponentially small number of points $(x,y)$ around $x=y$ could significantly contribute to the expectation of $K(x,y)$. Specifically, the number of points near $x,y$ is $\mathcal{O}(2^n)$, whereas the total number of points $(x,y)$ is $\mathcal{O}(2^{2n})$. Consequently, in dense distributions where amplitudes are nearly uniform across all points $x$ (compared to the sparse distribution),the amplitude $p(x)$ at a given point $x$ would drop exponentially as problem size increases, finally leading to the exponentially decay of MMD loss. To mitigate this issue, we can take the logarithm of the original MMD loss function, which helps counterbalance the exponential decay and enhances the stability of the loss function value.

\begin{figure*}[t]
  \centering
  \begin{subfigure}{0.45\textwidth}
    \includegraphics[width=\textwidth]{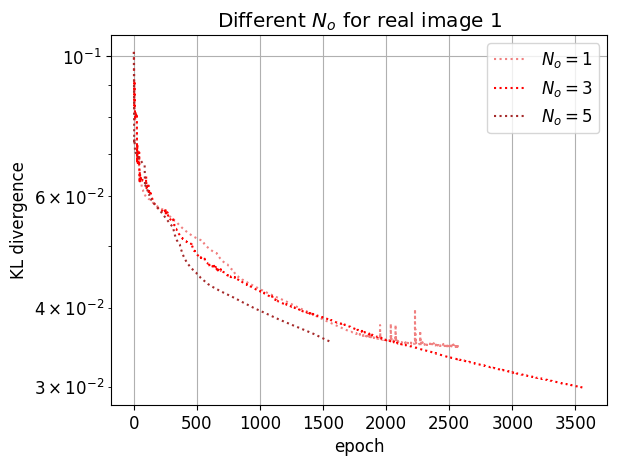}
    \caption{}
    \label{fig:image1}
  \end{subfigure}
  \hfill
  \begin{subfigure}{0.425\textwidth}
    \includegraphics[width=\textwidth]{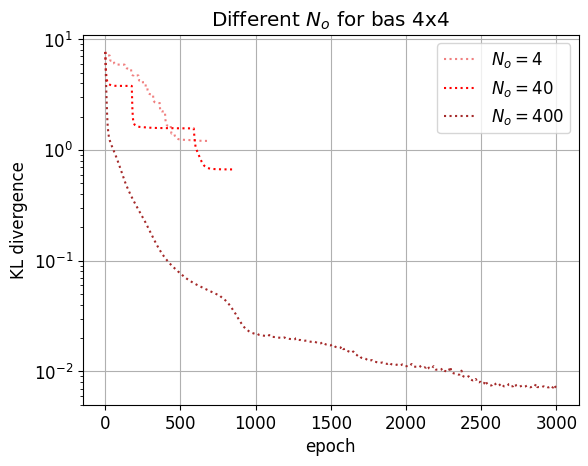}
    \caption{}
    \label{fig:image2}
  \end{subfigure}
  \caption{The effect of varying the number of operators, $N_o$, on the performance of ACLBM with different datasets. (a) For the densely distributed real image 1 dataset, a higher $N_o$ leads to faster convergence with fewer iterations required: 250 iterations for $N_o=1$, 204 for $N_o=3$, and 124 for $N_o=5$, while using 250, 612, and 620 parameters, respectively. (b) For the sparser BAS $4\times 4$ dataset, the number of iterations to convergence decreases significantly with an increase in $N_o$: 16 iterations for $N_o=4$, 5 for $N_o=40$, and only 1 for $N_o=400$, corresponding to a parameter count of 64, 200, and 400, respectively.}
  \label{fig:different_No}
\end{figure*}

\section{Discussions}
In this section, we first examine factors that influence the stability and convergence speed of our model. These factors include the number of operators selected during each epoch (Section~\ref{subsec:number_of_ops}) and strategies for adjusting the learning rate (Section~\ref{subsec:adjustable_lr}). Secondly, we explore methods to reduce the number of operators in the operator pool (Section~\ref{subsec:op_pool_reduction}), aiming to decrease the number of measurements required in each epoch. Additionally, we conduct a numerical analysis to assess the impact of this reduction on our model's performance. Furthermore, we implement the image remapping technique proposed in Ref.~\cite{zhou2023hybrid} to determine whether ACLBM can efficiently learn such transformed distributions (Section~\ref{subsec:image_remapping}). We also study the effects of varying discretization precision on distributions (Section~\ref{subsec:different_disc_precision}) to assess whether increasing the number of auxiliary qubits impacts the computational resources required by ACLBM. Lastly, we discuss two potential applications of our model: the first involves approximating the oracle $\text{PREP}$ used in block encoding (Section~\ref{section:Hamiltonian_Simulation}), and the second focuses on preparing the log-normal state for Monte Carlo pricing (Section~\ref{subsec:MC_pricing}).

\subsection{Number of operators.}
\label{subsec:number_of_ops}
We explore the impact of varying the number of operators, $N_o$, per iteration on the performance of the Adaptive Circuit Learning of Born Machine (ACLBM). As illustrated in Fig.~\ref{fig:different_No}, when applied to a dense distribution (specifically, the real image 1 dataset in our study), a moderate increase in $N_o$ above 1 slightly enhances both performance and convergence rate. Notably, performance appears to saturate at $N_o=3$. While the $N_o=5$ setting achieves faster convergence and requires slightly more parameters than the $N_o=3$ scenario, the overall converged KL divergence for these two settings is comparable.  In the case of sparse distributions, exemplified by the BAS $4\times 4$ dataset, a larger $N_o$ is essential to ensure convergence. With $N_o=4$, convergence is notably sluggish, and the final KL divergence hovers around 1.2. However, as $N_o$ increases to 40 or even 400, we see a marked acceleration in convergence and an improved KL divergence. This effect likely stems from the BAS dataset's inherent need for long-range entanglement, where a smaller $N_o$ provides insufficient spread of entanglement. This can cause the optimization process to settle into local minima during the first few steps, preventing the model from accurately representing the BAS distribution.

\begin{figure}
    \centering
    \includegraphics[width=\linewidth]{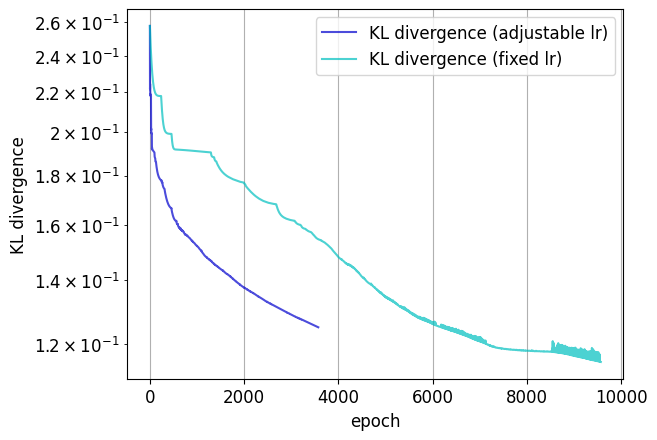}
    \caption{Comparison of KL divergence with adjustable and fixed learning rates. The adjustable learning rate dynamically adapts based on the gradient norm, leading to smoother and more stable convergence, while the fixed learning rate results in oscillations and instability as the model nears convergence.}
    \label{fig:adjustable_lr}
\end{figure}

\subsection{Adjustable learning rate.}
\label{subsec:adjustable_lr}
To ensure stability and rapid convergence of our model, we adopt an adjustable learning rate, reassigning it a new value at the start of each epoch. This approach prevents fluctuations in the loss function, which commonly occur when a constant learning rate is too high as the model approaches convergence. Specifically, we base our learning rate on the norm of the gradient of selected operators, setting it as follows:

\begin{equation}
\text{lr}=\alpha \cdot \frac{\|\bm{g}\|_2}{\sqrt{N_o}}.
\end{equation}

Here, $\alpha$ is a scaling factor, generally chosen between 0.05 to 0.5, and $\|\bm{g}\|_2$ is the 2-norm of the gradient vector $\bm{g}$. This strategy allows the learning rate to adapt dynamically, promoting fast and stable convergence. 

As illustrated in Fig.~\ref{fig:adjustable_lr}, the model using a fixed learning rate exhibits oscillations in the loss function, often due to an excessively large learning rate as the model approaches convergence. This instability is particularly evident when a new operator is appended, causing the loss function to increase at the beginning of each iteration. In contrast, the adjustable learning rate effectively mitigates this issue, leading to a smoother and more consistent reduction in loss.

\begin{figure}[t]
  \centering
  \begin{subfigure}{0.45\textwidth}
    \includegraphics[width=\textwidth]{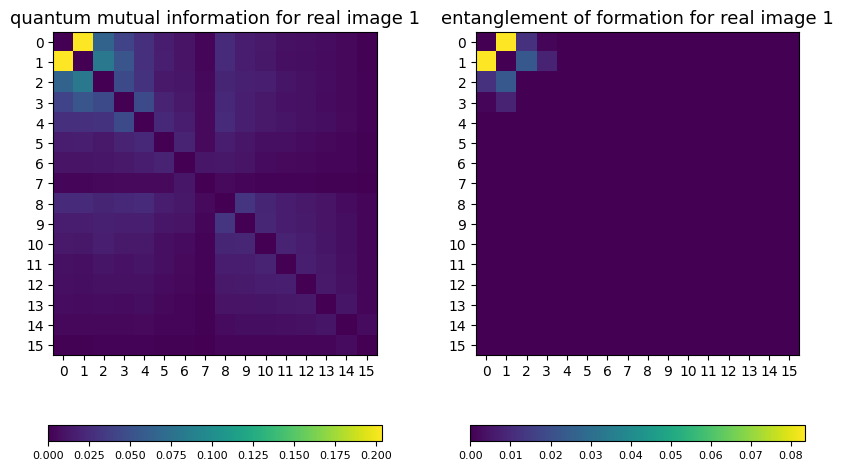}
    \caption{}
    \label{fig:entangle image1}
  \end{subfigure}
  \hfill
  \begin{subfigure}{0.45\textwidth}
    \includegraphics[width=\textwidth]{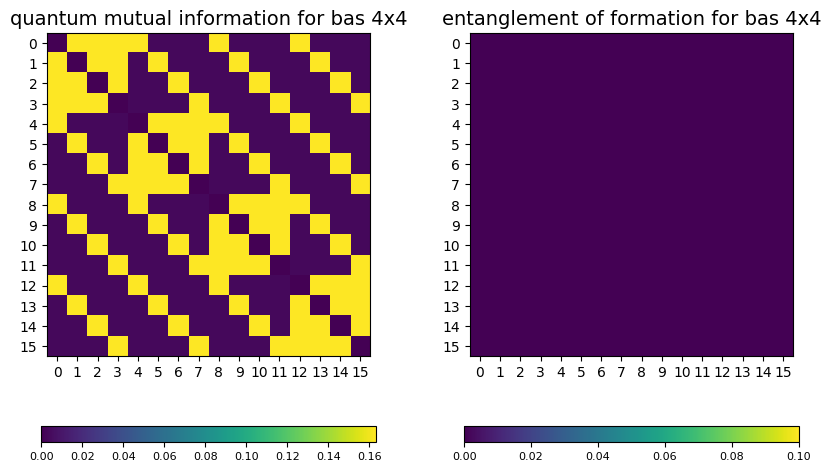}
    \caption{}
    \label{fig:entangle image2}
  \end{subfigure}
  \caption{Quantum mutual information and entanglement of formation for two different datasets. The axes correspond to qubit indices, and the correlation is evaluated on two-qubit subsystems of the target state $|\Psi^\star\rangle$ in Eq.~(\ref{eq:target_state}). The color scale indicates the strength of the correlation, with brighter colors denoting stronger correlations and darker colors indicating weaker ones.
  Panel (a) shows the entanglement correlation for a real image dataset, where both quantum mutual information and entanglement of formation identify pairs of qubits with strong correlations. Quantum mutual information generally indicates a smoother transition than entanglement of formation. 
  Panel (b) displays the entanglement correlation for the BAS $4\times 4$ dataset. Here, only quantum mutual information reflects the two-qubit correlation, whereas entanglement of formation registers zero correlation, suggesting that the BAS dataset may exhibit long-range entanglement, rendering two-qubit subsystem entanglement of formation immeasurable.}
  \label{fig:correlation}
\end{figure}

\begin{figure}[ht]
    \centering
    \begin{subfigure}{0.45\textwidth}
        \centering
        \includegraphics[width=\textwidth]{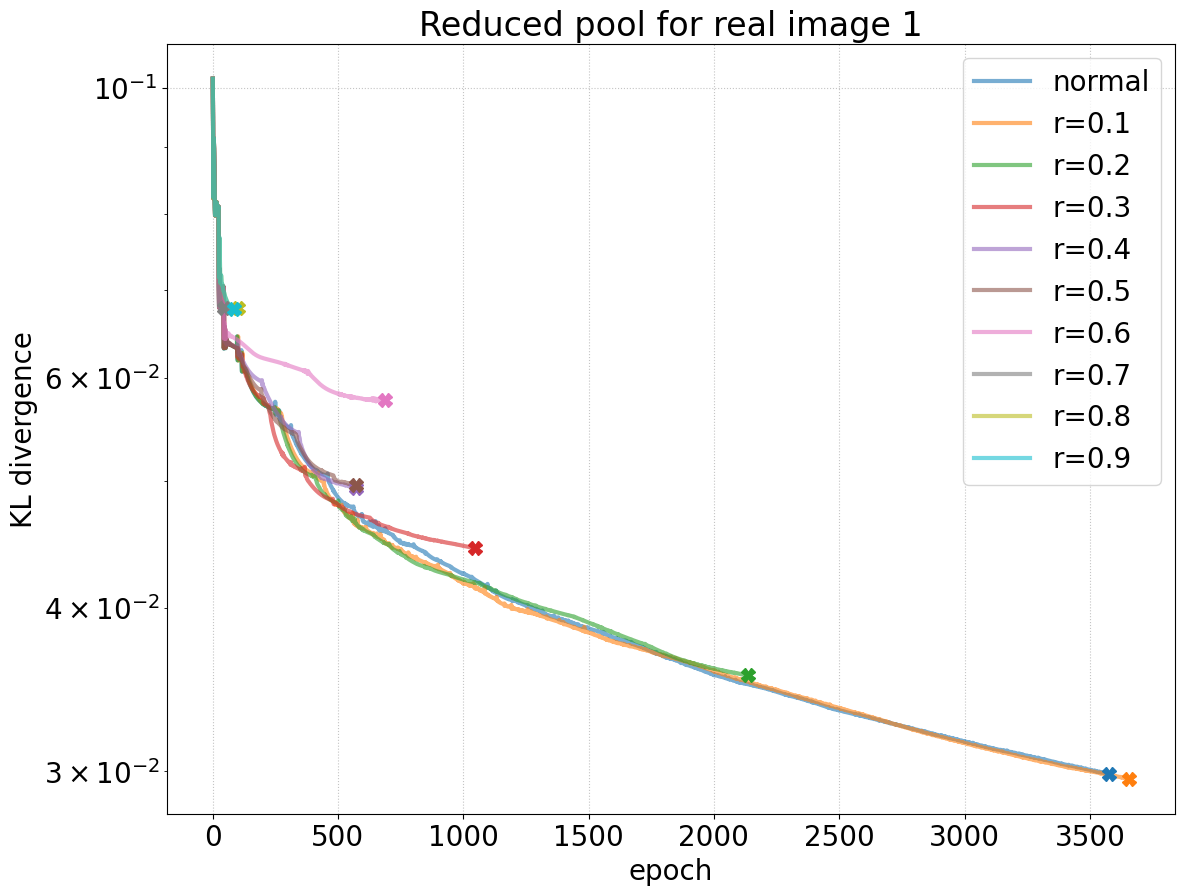}
        \caption{}
        \label{fig:reduce pool for image 1}
    \end{subfigure}
    \begin{subfigure}{0.45\textwidth}
        \includegraphics[width=\textwidth]{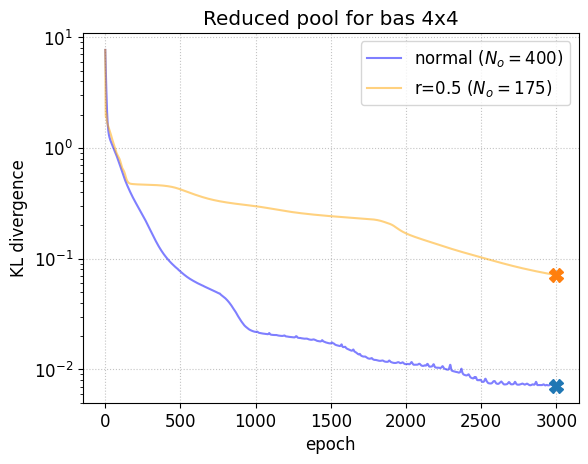}
        \caption{}
        \label{fig:reduce pool for bas 4x4}
    \end{subfigure}
    \caption{Effect of operator pool reduction for different datasets. (a) For the real image 1 dataset, reduction rates from 0 to 0.9 were tested. Results indicate that low reduction rates do not significantly impact the convergence of KL divergence. However, higher reduction rates result in a faster performance deterioration. (b) For the BAS $4\times 4$ dataset, a reduction rate of 0.5 was selected based on the sharp transition in mutual information observed in Fig.~\ref{fig:correlation}, targeting qubit pairs with strong correlations. Post-reduction, the reduced number of operators with significant gradients decrease $N_o$ from 400 to 175. This adjustment led to a decrease in performance compared to normal one.}
    \label{fig:pool reduction}
\end{figure}

\begin{figure*}
    \centering
    \includegraphics[width=\textwidth]{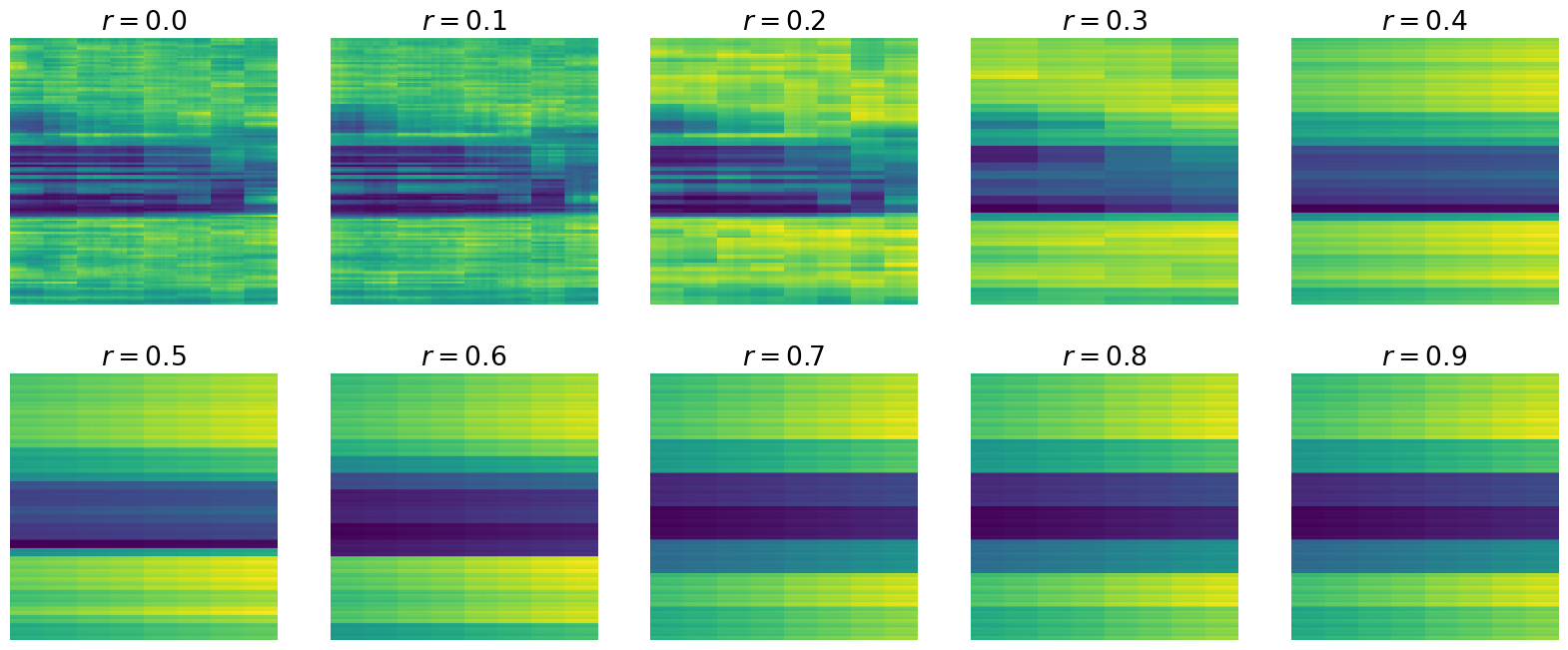}
    \caption{Images generated by ACLBM with varying operator pool reduction rates.}
    \label{fig:reduce_rate_image_array}
\end{figure*}

\subsection{Operator Pool Reduction.} 
\label{subsec:op_pool_reduction}
Our algorithm, as it currently stands, evaluates the gradients of $\mathcal{O}(n^2)$ operators at each iteration. Yet, in certain distributions, specific qubit pairs may contribute more significantly to the overall quantum correlation than others. In such cases, we can streamline the operator pool by retaining only those operators that act on highly correlated qubit pairs, thereby eliminating the less contributory ones. There are numerous methods to assess the correlation between two qubits in a subsystem. In this study, we employ quantum mutual information and the entanglement of formation (detailed discussion can be found in Appendix~\ref{Appendix:Quantum Mutual Information and Entanglement of Formation}) — both of which have well-defined expressions in the two-qubit scenario—to discern correlation patterns, here we take real image 1 and BAS $4\times 4$ dataset for example, the correlation between different qubit pairs can be seen in Fig.~\ref{fig:correlation}.

Fig.~\ref{fig:pool reduction} illustrates the effects of reducing the operator pool within our adaptive circuit learning framework. We define the reduction rate \( r \) as the threshold for selecting two-qubit operators based on their correlation relative to the maximum observed correlation in the system:

\begin{equation}
\label{eq:reduced_pool}
    \mathcal{P}_r = \left\{ G_{ij} \in \mathcal{P} \, \middle| \, I(q_i, q_j) \geq r \cdot I_{\text{max}} \right\} \cup \left\{ G_i \in \mathcal{P} \right\},
\end{equation}
\begin{equation}
    I_{\text{max}} = \max_{ij} I(q_i, q_j),
\end{equation}
\\
where $\mathcal{P}_r$ represents the reduced operator pool, $\mathcal{P}$ denotes the original full pool, and $ I(q_i, q_j) $ is the quantum mutual information between qubits indexed $i$ and $j$. Here, $G_{ij}$ symbolizes the two-qubit gates and  $G_i$ the single-qubit gates. This methodical reduction of the operator pool aims to retain only the most impactful gates, thereby streamlining the learning process while preserving the ability to capture essential quantum correlations. Fig.~\ref{fig:reduce pool for image 1} presents experiments with reduction rates ranging from 0 to 0.9 for the real image 1 dataset. We observe that low reduction rates do not adversely affect the convergence of KL divergence and yield comparable performance, implying that moderate operator pruning is viable without significantly deteriorating the algorithm's performance. However, higher reduction rates result in a rapid decline in effectiveness. This decline may be attributed to the exclusion of highly correlated qubit pairs, which diminishes the model's ability to accurately capture the correlations within the distribution. The resulting image of different reduction rates can be found in Fig.~\ref{fig:reduce_rate_image_array}

For the BAS $4\times 4$ dataset, as shown in Fig.~\ref{fig:entangle image2}, we observe a stark transition in quantum mutual information. In response, we implemented a reduction rate of 0.5 to identify operator pairs with heightened correlations. After reducing the operator pool, we selected the top $N_o=175$ operators with the most significant gradient values. The numerical results in Fig.~\ref{fig:reduce pool for bas 4x4}, indicate that, while the performance with the reduced operator pool shows a slightly higher converged KL divergence, it also requires fewer resources and shallower circuit depth. These findings highlight the nuanced trade-offs of operator pool reduction and emphasize the importance of strategic operator selection based on quantum correlation metrics to maintain robust algorithmic performance.

\begin{figure}
    \centering
    \begin{subfigure}{0.15\textwidth}
        \centering
        \includegraphics[width=\textwidth]{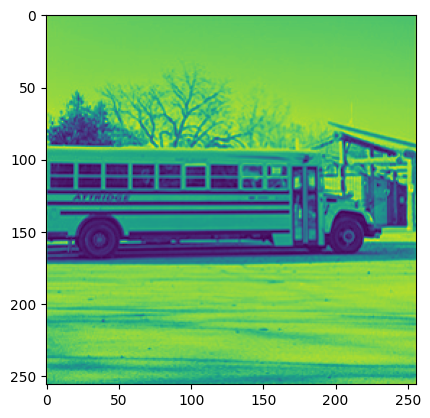}
    \end{subfigure}
    \begin{subfigure}{0.15\textwidth}
        \centering
        \includegraphics[width=\textwidth]{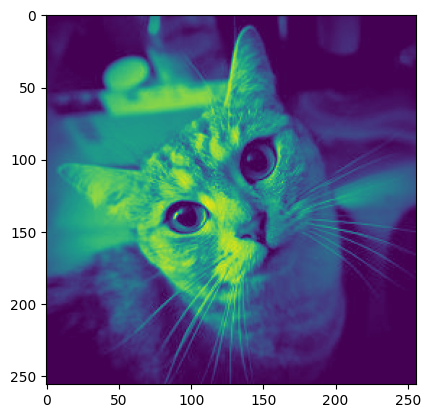}
    \end{subfigure}
    \begin{subfigure}{0.15\textwidth}
        \centering
        \includegraphics[width=\textwidth]{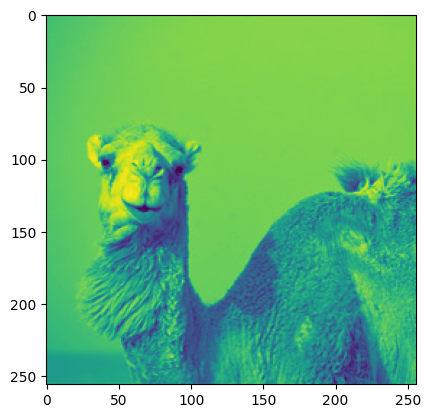}
    \end{subfigure}
    \begin{subfigure}{0.15\textwidth}
        \centering
        \includegraphics[width=\textwidth]{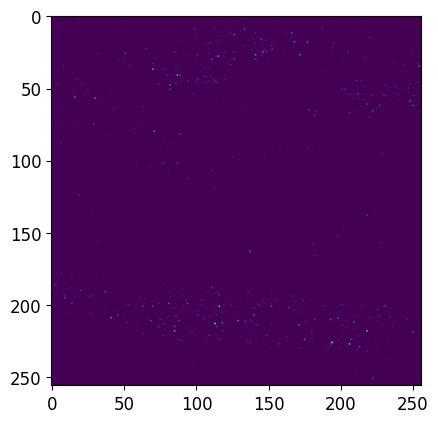}
    \end{subfigure}
    \begin{subfigure}{0.15\textwidth}
        \centering
        \includegraphics[width=\textwidth]{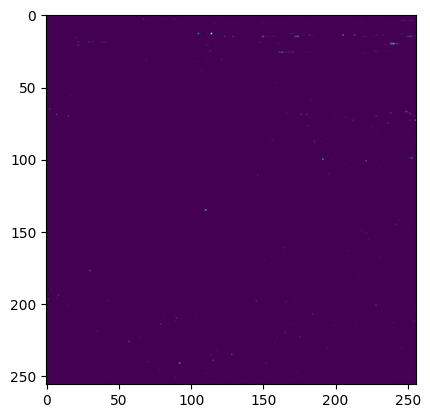}
    \end{subfigure}
    \begin{subfigure}{0.15\textwidth}
        \centering
        \includegraphics[width=\textwidth]{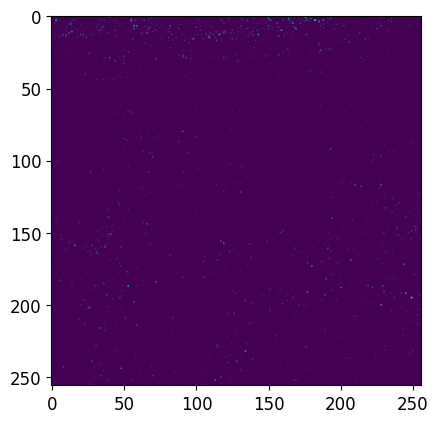}
    \end{subfigure}
    \begin{subfigure}{0.15\textwidth}
        \centering
        \includegraphics[width=\textwidth]{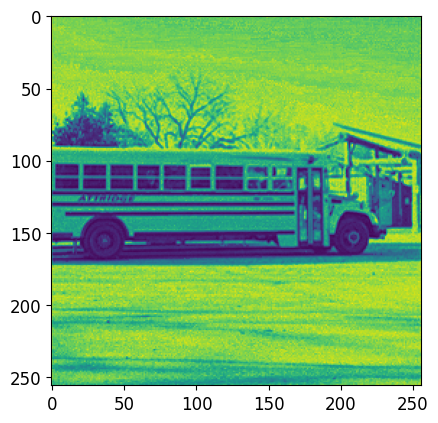}
    \end{subfigure}
    \begin{subfigure}{0.15\textwidth}
        \centering
        \includegraphics[width=\textwidth]{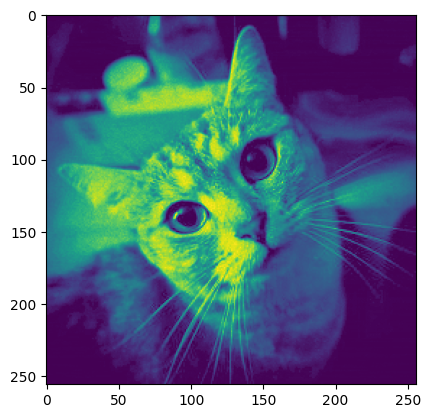}
    \end{subfigure}
    \begin{subfigure}{0.15\textwidth}
        \centering
        \includegraphics[width=\textwidth]{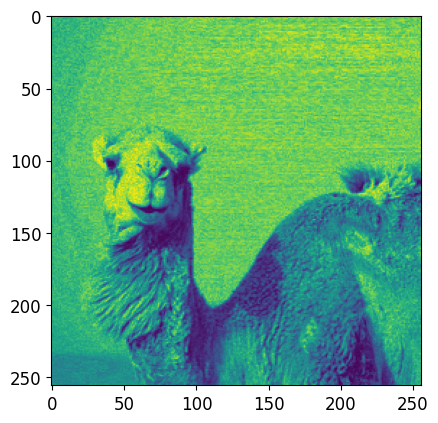}
    \end{subfigure}
    \begin{subfigure}{0.15\textwidth}
        \centering
        \includegraphics[width=\textwidth]{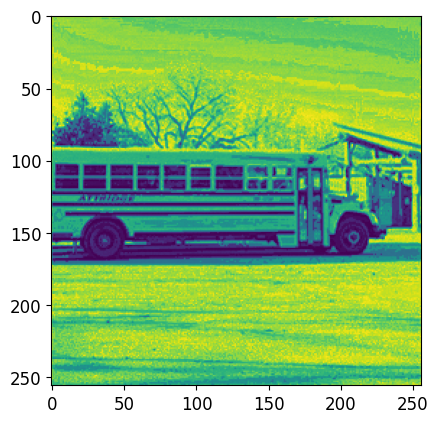}
    \end{subfigure}
    \begin{subfigure}{0.15\textwidth}
        \centering
        \includegraphics[width=\textwidth]{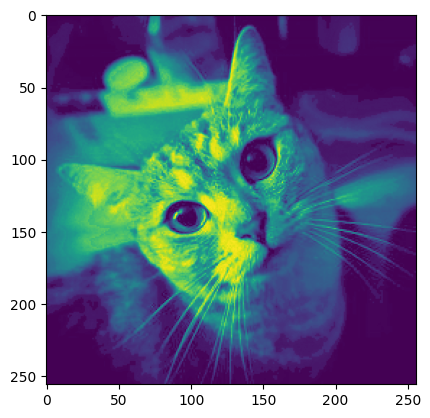}
    \end{subfigure}
    \begin{subfigure}{0.15\textwidth}
        \centering
        \includegraphics[width=\textwidth]{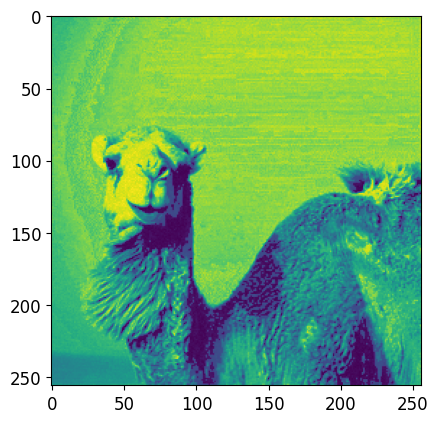}
    \end{subfigure}
    \begin{subfigure}{0.15\textwidth}
        \centering
        \includegraphics[width=\textwidth]{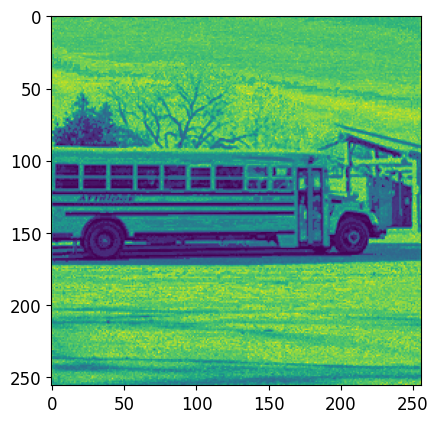}
    \end{subfigure}
    \begin{subfigure}{0.15\textwidth}
        \centering
        \includegraphics[width=\textwidth]{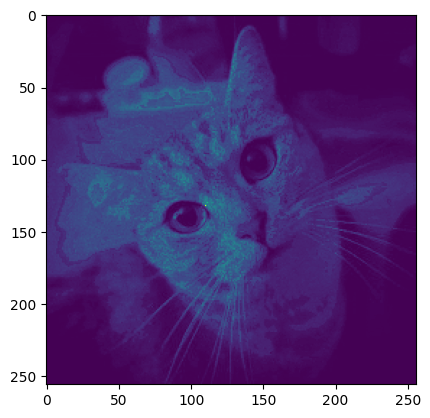}
    \end{subfigure}
    \begin{subfigure}{0.15\textwidth}
        \centering
        \includegraphics[width=\textwidth]{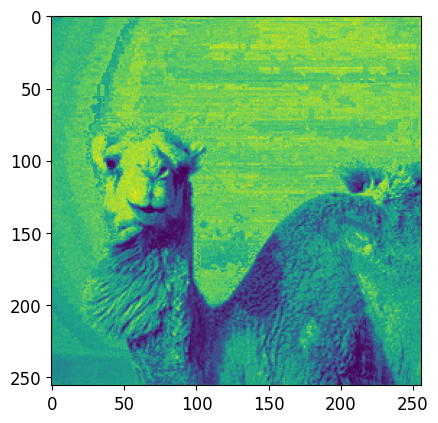}
    \end{subfigure}
    \begin{subfigure}{0.15\textwidth}
        \centering
        \includegraphics[width=\textwidth]{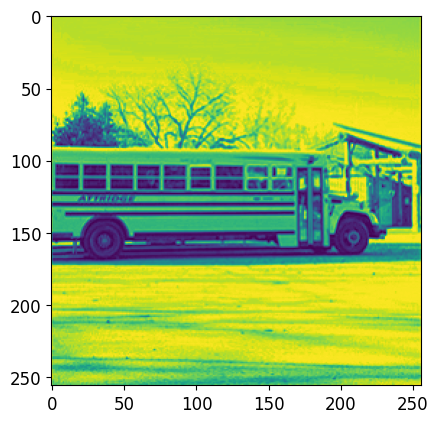}
    \end{subfigure}
    \begin{subfigure}{0.15\textwidth}
        \centering
        \includegraphics[width=\textwidth]{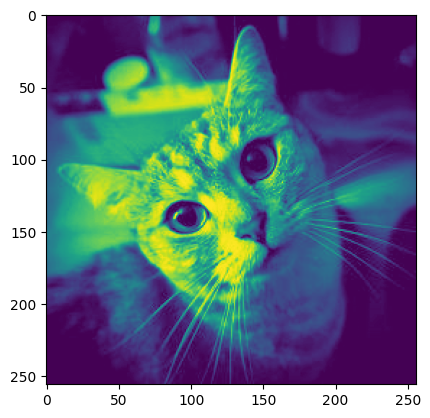}
    \end{subfigure}
    \begin{subfigure}{0.15\textwidth}
        \centering
        \includegraphics[width=\textwidth]{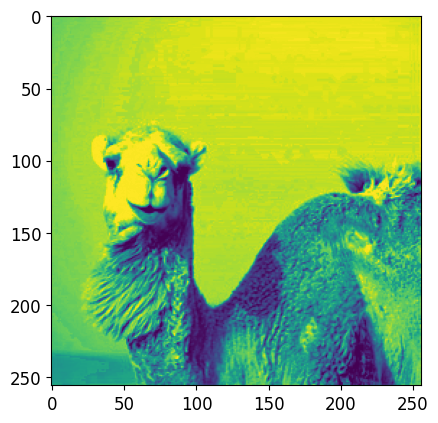}
    \end{subfigure}
    \caption{Reconstructed images generated by different models using the image remapping technique. The top row displays the original images 1, 2, and 3, while each subsequent row shows results from a different model, listed from top to bottom as QGAN, QCBM, DDQCL, MPS, and ACLBM, respectively.}
    \label{fig:image_remapping}
\end{figure}

\subsection{Image remapping}
\label{subsec:image_remapping}

In Ref.~\cite{zhou2023hybrid}, the author proposes a method to remap image pixels to make the distribution easier to learn. This approach involves sorting the pixel intensities in ascending order, which transforms the original distribution into a much smoother one. The smoother distribution is generally easier for learning algorithms, such as the proposed ACLBM, to handle effectively. Based on previous conjectures and numerical results, we suggest that smooth functions are more accessible for ACLBM to learn, making this remapping technique particularly advantageous for enhancing performance.

We conducted numerical experiments using real image data, and the results were exceptionally positive. The reconstructed images in Fig.~\ref{fig:image_remapping} showed much higher resolution and clarity compared to those trained directly on the original image distribution. It is important to note that the images are reconstructed by reversing the sorting process. Specifically, if the sorting operation corresponds to a permutation function $\pi: \{1,...,2^n\}\to\{1,...,2^n\}$, then we apply the inverse permutation function $\pi^{-1}$  to the output distribution to restore the original image distribution. This process ensures that the transformed data can be learned efficiently, while the final output still accurately represents the original image.

\begin{figure}
    \centering
    \includegraphics[width=\linewidth]{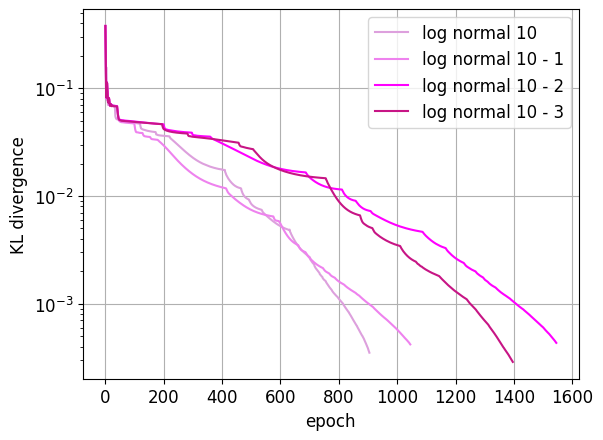}
    \caption{KL divergence convergence of ACLBM on log-normal distributions with varying discretization precisions, corresponding to 1, 2, and 3 auxiliary qubits. The results demonstrate that increasing discretization precision has minimal impact on convergence and complexity, with all configurations reaching similar KL divergence values.}
    \label{fig:discretized_precision}
\end{figure}

\subsection{Different discretization precision of continuous probability}
\label{subsec:different_disc_precision}
In this subsection, we investigate the performance of ACLBM on a discretized continuous distribution with varying discretization precision. We use the discretization method described in Eq.~(\ref{eq:discretized_cont_prob}) on a log-normal distribution with $\mu=5.5, \sigma=0.9$ considering distributions with 1, 2, and 3 auxiliary qubits. The numerical results, as shown in Fig.~\ref{fig:discretized_precision}, indicate that for all discretizations with different precision levels, the model converges to nearly the same KL divergence and requires approximately the same resources. This suggests that increasing discretization precision does not significantly increase complexity. These findings align with the argument in Ref.~\cite{holmes2020efficient, garcia2021quantum} that entanglement entropy grows very slowly as the number of auxiliary qubits increases.

\begin{figure*}[th]
\subcaptionbox{}[0.45\linewidth][c]{
\begin{quantikz}
    \lstick{$\ket{0}_a$} & \qwbundle{a} & \gate{\text{PREP}} & \gate[2,disable auto height]{\text{SEL}} & \gate{\text{PREP}^\dagger} && \\
    \lstick{$|\psi\rangle_s$} & \qwbundle{s} &&&&&
\end{quantikz}
}
\subcaptionbox{}[0.5\linewidth][c]{
\begin{quantikz}
    &  \gate[2,disable auto height]{\text{SEL}}&
    \\
    &  &
\end{quantikz}
=
\begin{quantikz}
    & &\octrl{2}&\octrl{2}&\ctrl{2}&\ctrl{2}& \\
    & &\ocontrol{}&\control{}&\ocontrol{}&\control{}& \\ 
    & \qwbundle{s} &\gate{V_1}&\gate{V_2}&\gate{V_3}&\gate{V_4}&
\end{quantikz}
}
\caption{Block encoding circuit structure. (a) depicts the general circuit for implementing the linear combination of unitaries (LCU). Two oracles are utilized: PREP, which prepares an arbitrary superposition as outlined in Eq.~(\ref{eq:BE_oracle}), and SEL, which governs the multi-qubit control over the distinct terms $V_j$ for $j \in \{1, \ldots, d\}$. (b) illustrates a simplified case for $d = 4$ with two ancilla qubits, showcasing the practical realization of the SEL oracle.
\label{fig:block_encoding}}
\end{figure*}

\subsection{Applicability for Hamiltonian Simulation}
\label{section:Hamiltonian_Simulation}
A substantial application of our proposed framework is in the construction of the approximated oracle within fault-tolerant quantum algorithms. Recent advancements in quantum computing have highlighted the prolific adoption of the ``block encoding" technique, which embeds a matrix within the top left block of a unitary operator. This embedding is primarily achieved through a method known as the ``linear combination of unitaries" (LCU)~\cite{childs2012hamiltonian}. This approach plays a crucial role in various quantum algorithms, including Qubitization~\cite{low2019hamiltonian} and Quantum Singular Value Transformation~\cite{gilyen2019quantum}. It effectively combines the principles of quantum signal processing with block encoding, thereby enabling polynomial transformations on block-encoded operators.

The essence of block encoding is relatively straightforward. It involves embedding a target operator $H$ within the top-left block of a unitary operator $U$:
\begin{equation}
    U=\begin{pmatrix}
        H / \alpha & \cdot \\ 
        \cdot & \cdot
    \end{pmatrix}.
\end{equation}
The normalization $\alpha$ is to ensure that the spectral norm of $H/\alpha$ is less than or equal to 1. This representation can be realized as follows: Initially, a Hermitian matrix $H$ is expressed as a linear combination of unitaries. This is directly related to practical scenarios where the fermionic Hamiltonian is mapped to Pauli operators:
\begin{equation}
\label{eq:block_encoding}
    H=\sum_{j=1}^d \alpha_j V_j, \hspace{0.4cm} \Vert H \Vert \leq \alpha = \sum_{j=1}^{d} \vert \alpha_j \vert, 
\end{equation}
Subsequently, we utilize two oracles, $\text{PREP}$ and $\text{SEL}$:
\begin{align}
\label{eq:BE_oracle}
    &\text{PREP}|0\rangle_a=\sum_{j=1}^{d}\sqrt{\frac{\alpha_j}{\alpha}}|j\rangle_a, \hspace{0.4cm} 
    \nonumber
    \\
    &\text{SEL}=\sum_{j=1}^{d}|j\rangle\langle j|_a\otimes V_j.
\end{align}
Here the subscript $a$ denotes the ancilla register, with this two oracle we can realize the block encoding 
\begin{align}
    &(\langle 0|_a\otimes \mathbbm{1}_s)\text{PREP}^\dagger \cdot \text{SEL} \cdot \text{PREP} (|0\rangle_a \otimes \mathbbm{1}_s) 
    \nonumber \\
    &=\frac{1}{\alpha}\sum_{j=1}^{d}\alpha_jV_j=H/\alpha,
\end{align}
the detailed circuit structure is shown in Fig.~\ref{fig:block_encoding}.  With this block-encoded form, we can apply quantum signal processing with the qubitization technique to polynomially transform this operator and implement the time evolution operator $e^{-iHt}$ up to error $\epsilon$ (measured in spectral norm).

However, while the oracle $\text{SEL}$ is straightforward to implement (simply by using $d$ controlled-$V_j$ operations), implementing the oracle $\text{PREP}$ might not be trivial. In realistic electronic structure problems, the number of terms could scale as $d = \mathcal{O}(n^4)$, where $n$ represents the number of spin orbitals. Our algorithm can serve as a method to find an approximated gate compilation, denoted as $\text{PREP}'$, of the oracle $\text{PREP}$.

Suppose that $\text{PREP}'$ prepares a superposition state
\begin{equation}
\text{PREP}'|0\rangle = \sum_{j=1}^{d}e^{i\phi_j}\sqrt{q_j}|j\rangle,
\end{equation}
where the KL divergence between the two distributions $\{q_j\}_{j=1}^{d}$ and $\{\alpha_j/\alpha\}_{j=1}^{d}$ is a small quantity $\delta$. Then, the approximated oracle block encodes an approximated Hamiltonian:
\begin{equation}
H' = \sum_{j=1}^{d}q_jV_j,
\end{equation}
The error of this approximation, compared to the true time evolution operator, scales as (a detailed proof can be found in Appendix~\ref{Appendix:Error_bound_time_evolution}):
\begin{equation}
\Vert e^{-iHt} - e^{-iH'\alpha t} \Vert \leq \sqrt{2\delta}\alpha t.
\end{equation}

Hence, if we want to approximate the time evolution operator up to $\epsilon$ error, the requirement of the precision of KL divergence can be obtained:
\begin{equation}
    \delta \leq \frac{\epsilon^2}{2\alpha^2t^2}.
\end{equation}

\begin{figure}[t]
    \includegraphics[width=0.45\textwidth]{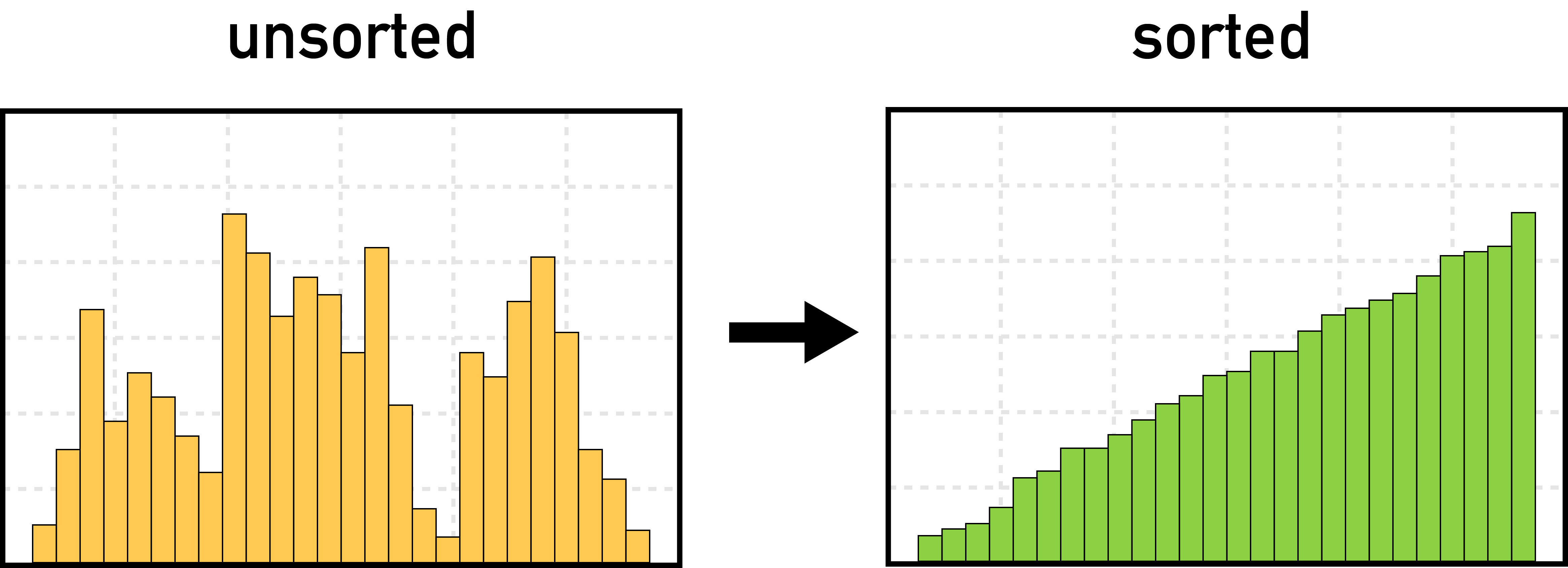}
    \caption{This figure graphically illustrates the unstructured distribution of coefficients arising from the decomposition within a linear combination of unitaries (LCU). The left panel displays the original distribution, which is typically derived from the coefficients of a Pauli decomposition of a Hamiltonian, while the right panel showcases the sorted arrangement. This rearrangement corresponds to a permutation $\pi: S\to S, S=\{1,...,d\}$. We relabel the indices through this permutation as  $\alpha_j \mapsto \alpha_{\pi^{-1}(j)}$, $V_{j}\mapsto V_{\pi^{-1}(j)}$.} 
    \label{fig:sorted_dist}
\end{figure}

It is worth noting that in some cases—and likely most cases within quantum chemistry—the desired coefficients $\alpha_j / \alpha$ for $j\in\{1,...,d\}$ might be highly unstructured. For example, the one electron integral $h_{pq}$ and two electron integral $h_{pqrs}$, where $p,q,r,s$ denote the labels of spin orbitals, depend heavily on the chosen basis sets and the geometry of the molecules, and thus often lack an obvious pattern. According to our previous numerical results, generating approximate forms of such distributions presents a significant challenge. However, it can be observed that in Eq.~(\ref{eq:block_encoding}), the decomposition of the operator $H$ remains invariant under any arbitrary permutation of the indices $j$. Therefore, we can exploit this degree of freedom to impose more structure on the coefficient $\alpha_j/\alpha$. This restructuring can be achieved by sorting the values of $|\alpha_j|$ in ascending order, followed by a corresponding relabeling of the indices in Eq.~(\ref{eq:block_encoding}). A simple example can be found in Fig.~\ref{fig:sorted_dist}.

It is also important to note that we currently consider only cost metrics relevant to the NISQ era, such as the total number of gates and overall circuit depth. However, in the context of fault-tolerant quantum computation, these cost metrics differ significantly. For instance, single-qubit rotations must be synthesized from fault-tolerant $T$-gates, and implementing arbitrary single-qubit rotations may require a substantial number of $T$-gates. It remains uncertain whether ACLBM retains its advantages under these conditions, and further numerical studies are needed to assess its performance with fault-tolerant cost metrics.

\subsection{Monte Carlo Pricing in Quantum Finance}
\label{subsec:MC_pricing}
Another potential application lies in estimating the expected payoff in options pricing. The standard financial model for calculating options pricing is the Black-Scholes-Merton (BSM) model~\cite{black1973pricing,merton1973theory}. It assumes that the stock price follows a geometric Brownian motion. The stochastic differential equation that defines the price dynamics of the risky asset $S_t$ is:
\begin{equation}
\label{eq:GBM}
    dS_t=S_t\mu dt + S_t\sigma dW_t,
\end{equation}
where $\mu$ is the drift (average return of the stock), $\sigma$ is the volatility (standard deviation of the stock's returns), and $W_t$ is the Wiener process satisfying some properties: \\
(1) $W_0=0$, \\
(2) $W$ has continous path, \\
(3) $(W_{t+u}-W_{t}) \perp W_{s}$ for all $u>0, s<t$, \\
(4) $W_{t+u}-W_{t}\sim\mathcal{N}(0,u)$.\\ 
Solving Eq.~(\ref{eq:GBM}), we arrive at the expression for $S_t$:
\begin{equation}
    S_t = S_0 e^{\sigma W_t + (\mu-\sigma^2/2)t},
\end{equation}
where at any given time $t=T$, $S_T$ follows a log normal distribution.

An important market assumption in the BSM model is the absence of arbitrage opportunities, meaning riskless profit cannot be made. This is mathematically characterized as:
\begin{equation}
    S_0 = e^{-rT} \mathbb{E}_{W_T}[S_T],
\end{equation}
where $e^{-rT}$ is the discounting factor,  determining the present value of future cash flows. To satisfy this assumption, we simply substitute the drift $\mu$ with the risk-free rate $r$.

One task of great interest is to estimate the price of options, such as the European call option. The payoff function $f(S_T)$ is given by
\begin{equation}
\label{eq:European_call_option}
    f(S_T) = \max\{0, S_T-K\},
\end{equation}
where $K$ is the strike price and $T$ is the maturity date. The expectation value of $f(S_T)$ in Eq.~(\ref{eq:European_call_option}) can be analytically determined by the BSM model. However, for more complex payoff functions, an analytical expression may not exist. In such cases, Monte Carlo simulation is required to estimate the pricing.

It is well established that quantum computing provides a nearly quadratic complexity improvement in Monte Carlo tasks. This improvement is achieved by reformulating Monte Carlo simulation into an amplitude estimation problem. Initially, we require a data loading oracle $\mathcal{A}$ and a control rotation oracle $\mathcal{R}$:
\begin{align}
    \mathcal{A}|0^{n}\rangle &=\sum_{x=0}^{2^n-1} \sqrt{p(x)} |x\rangle, \\ 
    \mathcal{R}|x\rangle|0\rangle &=|x\rangle[\sqrt{1-f(x)}|0\rangle+\sqrt{f(x)}|1\rangle].
\end{align}
Applying these two oracle to $|0^{n+1}\rangle$:
\begin{align}
    |\chi\rangle &\coloneqq \mathcal{R}(\mathcal{A}\otimes I)|0^{n+1}\rangle 
    \nonumber
    \\ &= \sum_{x=0}^{2^n-1} \sqrt{p(x)} [\sqrt{1-f(x)}|0\rangle+\sqrt{f(x)}|1\rangle].
\end{align}
We can immediately observe that the probability of measuring the state $|1\rangle$ on the ancilla qubit corresponds to the expectation value of function $f(x)$:
\begin{equation}
    \text{Tr}\big(|\chi\rangle\langle\chi|(I\otimes |1\rangle\langle 1|)\big) = \sum_{x=0}^{2^n-1}p(x)f(x).
\end{equation}
In the context of finance, $p(x)$  is chosen to follow a log-normal distribution, and $f(x)$ represents some complicated payoff function. However, it's important to note some caveats: proper truncation and discretization of $p(x)$ are necessary to encode the problem into a quantum circuit. Additionally, some normalization must be imposed to ensure that the payoff function, $f(x)$, is bounded within the interval $[0, 1]$~\cite{suzuki2020amplitude}.

To estimate the amplitude of the ancillary state $|1\rangle$, the well-known Quantum Amplitude Estimation (QAE) algorithm is employed. We will not delve into the details of the algorithm here but refer the reader to comprehensive references~\cite{brassard2002quantum,giovannetti2006quantum,suzuki2020amplitude} for further understanding. Utilizing QAE the number of queries to $f(x)$ required to estimate $\mathbb{E}[f(X)]$ up to error $\epsilon$ scales as $\mathcal{\tilde{O}}(\frac{\lambda}{\epsilon})$ nearly quadratically faster than classical Monte Carlo $\mathcal{O}(\frac{\lambda}{\epsilon})$, where $\lambda$ is the upper bound of the variance $\mathbb{V}[f(X)]\leq\lambda^2$. 

The error in the expectation value when using a learned distribution $q_{\bm{\theta}}$ instead of the true distribution $p(x)$ can be upper-bounded by $\sqrt{2\delta}\cdot\Vert f\Vert_2$, where $D_{\text{KL}}(p\Vert q_{\bm{\theta}})\leq\delta$ and $f_2$ is defined as $\left(\sum^{2^n-1}_{x}\vert f(x)\vert^2\right)^{1/2}$. The detailed proof of this error bound can be found in Appendix~\ref{Appendix:Error Bound of the Approximated Expectation Value}.

\section{Conclusions}

In conclusion, our study introduces the Adaptive Circuit Learning of Born Machine (ACLBM), a novel framework that significantly advances the capabilities of hybrid quantum-classical algorithms on the quantum data loading and state preparation problems. Leveraging the adaptability and precision of the ADAPT-VQE methodology, ACLBM showcases remarkable performance across various datasets, from simple probability distributions to the complex structures of real image data. This success underscores the framework's potential to solve the data loading problem with practical real world applications and suggests its integration into subroutines of more complicated quantum algorithms requiring to prepare the designate quantum states.

Our extensive numerical simulations showcase ACLBM's exceptional accuracy and efficiency in preparing quantum states with intricate correlations, outperforming other benchmark models. Furthermore, despite the inherent challenges, ACLBM successfully generates approximated amplitude embeddings of real-image data with human-recognizable features, a feat other models do not accomplish. The challenges encountered with QGAN~\cite{zoufal2019quantum} and QCBM~\cite{liu2018differentiable} in larger problem sizes emphasize the need for careful hyperparameter selection and the design of circuit architectures that can cope with the complexity of the data involved. Our findings also suggest that while the MMD loss function is suitable for sparse datasets, it may need to take logarithm in the dense dataset in order to counterbalance the exponential vanishing gradient problem.

Moreover, the potential applications of ACLBM in areas such as constructing oracle for Hamiltonian simulation and Monte Carlo pricing in quantum finance highlight the broad applicability of our work. By enabling efficient and accurate quantum data loading, our framework paves the way for realizing these complicated quantum algorithms, potentially advancing fields ranging from quantum chemistry to financial modeling.

Looking ahead, our study opens up several avenues for future exploration. One area of particular interest is the refinement of the adaptive circuit learning framework to further optimize parameter efficiency and convergence rates. Another promising direction is the investigation of different loss functions and their impact on the model's ability to approximate complex distributions more accurately. Additionally, our work offers compelling evidence that certain highly structured quantum states, which are characterized by a limited number of parameters, can be efficiently prepared using a relatively small number of gates. This raises the intriguing question of whether there is a deterministic algorithm that can directly acquire the gate compilation to prepare these states, instead of relying on an adaptive method to select them. It is also currently unknown what precision of amplitude embedding is necessary for the approximation to be meaningful. Is a very precise quantum state required, or are states that capture the most critical features sufficient? We expect that further theoretical work is required to answer these questions.

As quantum computing continues to advance, the algorithms and techniques developed for quantum data loading will play a crucial role in harnessing the full potential of quantum resources. Our work represents a simple methodology toward achieving this goal, providing a foundation for future research in the field of quantum algorithm and quantum machine learning.

\section*{Acknowledgements}
H.-C.~Cheng is supported by the Young Scholar Fellowship (Einstein Program) of the National Science and Technology Council, Taiwan (R.O.C.) under Grants No.~113-2119-M-007-006, No.~NSTC 113-2119-M-007-006, No.~NSTC 113-2119-M-001-006, No.~NSTC 113-2124-M-002-003, and No.~NSTC 113-2628-E-002-029, by the Yushan Young Scholar Program of the Ministry of Education, Taiwan (R.O.C.) under Grants No.~NTU-113V1904-5 and by the research project ``Pioneering Research in Forefront Quantum Computing, Learning and Engineering'' of National Taiwan University under Grant No.~NTU-CC-113L891605. H.-C.~Cheng acknowledges the support from the “Center for Advanced Computing and Imaging in Biomedicine (NTU-113L900702)” through The Featured Areas Research Center Program within the framework of the Higher Education Sprout Project by the Ministry of Education (MOE) in Taiwan.

\vspace{0.3cm}

\section*{Code Availability}
The code for ACLBM and the implementation of other benchmark models is available in the GitHub repository at \url{https://github.com/chuntse0514/Adaptive-Circuit-Learning-of-Born-Machine}.

\section*{Competing Interests}
The authors declare no competing interests.

\section*{Author Contributions}
C-T.L. proposed and implemented the ACLBM framework, and conducted the numerical experiments in comparison with previous models. H-C.C. provided insightful discussions on the error bounds and potential applications of ACLBM. All authors contributed to the final version of the manuscript.

\bibliographystyle{apsrev4-2}
\bibliography{references.bib}

\begin{widetext}
\appendix
\section{Zero gradient operators}
\label{Appendix:Zero gradient operators}
We establish that gradients vanish when initiating from a quantum state with solely real amplitudes and applying operators generated by an even number of Pauli $Y$ operators. This phenomenon can be inspected as follows: Introducing an operator into a circuit yields
\begin{align}
\frac{\partial q_{\bm{\theta}}(x)}{\partial \theta_i} &= \text{Tr}\left(|x\rangle\langle x| \frac{\partial}{\partial\theta_i}\left[U_{i}(\theta_i)\rho(\bm{\theta})U^\dagger_{i}(\theta_i)\right]\right)\Bigg|_{\theta_i=0} 
\nonumber
\\
&=-\frac{i}{2}\text{Tr}\left(|x\rangle\langle x| [P_i,\rho(\bm{\theta})]\right)
\nonumber
\\
&=-\frac{i}{2}\left(\langle x |P_i\rho(\bm{\theta})|x\rangle-\langle x|\rho(\bm{\theta})P_i |x\rangle\right) 
\nonumber
\\
&=-\frac{i}{2}\left(\langle x |P_i\rho(\bm{\theta})|x\rangle-\langle x |P_i\rho(\bm{\theta})|x\rangle^*\right) 
\nonumber
\\
&= 0,
\end{align}
where the last equality is deduced from the initial state $\rho(\bm{\theta})$ and the generator $P_i$ being real. Consequently, the absence of imaginary components results in a zero gradient due to the real nature of the elements involved.

\section{Gradient evaluation}
\label{Appendix:Gradient evaluation}
We illustrate that the gradient of a parameter can be explicitly computed with parameters-shift rule~\cite{mitarai2018quantum, schuld2019evaluating}, requiring three quantum circuit evaluations for each parameter. The derivative of the loss function $\mathcal{L}(\bm{\theta})$ with respect to a specific parameter $\theta_i$ is given by:
\begin{align}
\frac{\partial\mathcal{L}(\bm{\theta})}{\partial \theta_i} = -\sum_{x} \frac{p(x)}{q_{\bm{\theta}}(x)} \frac{\partial q_{\bm{\theta}}(x)}{\partial \theta_i}, \\
\frac{\partial\mathcal{L}(\bm{\theta})}{\partial \theta_i} =\frac{\sum_x\sqrt{\frac{p(x)}{q_{\bm{\theta}}(x)}}\frac{\partial q_{\bm{\theta}}(x)}{\partial \theta_i}}{2\sqrt{1-\langle \sqrt{p},\sqrt{q_{\bm{\theta}}}\rangle^2}}
\end{align}
where we introduce the following operators:
\begin{align}
\Pi_x(\bm{\theta}) &\equiv U^\dagger_\ell(\theta_\ell)\ldots U^\dagger_{i+1}(\theta_{i+1})|x\rangle\langle x|U_{i+1}(\theta_{i+1})\ldots U_{\ell}(\theta_{\ell}) 
\nonumber
\\
\rho(\bm{\theta})&\equiv U_{i-1}(\theta_{i-1})\ldots U_{1}(\theta_1)|0\rangle\langle0|U^\dagger_{1}(\theta_1)\ldots U^\dagger_{i-1}(\theta_{i-1}).
\end{align}
The derivative of $q_{\bm{\theta}}(x)$ with respect to $\theta_i$ can be expanded as
\begin{align}
\frac{\partial q_{\bm{\theta}}(x)}{\partial \theta_i} &= \text{Tr}\Biggl(\Pi_x(\bm{\theta}) \frac{\partial}{\partial\theta_i}\Bigl[U_{i}(\theta_i)\rho(\bm{\theta})U^\dagger_{i}(\theta_i)\Bigr]\Biggr) 
\nonumber
\\
&= -\frac{i}{2}\text{Tr}\Biggl(\Pi_x(\bm{\theta})U_i(\theta_i)[P_i, \rho(\bm{\theta})]U^\dagger_i(\theta_i) \Biggr) 
\nonumber
\\
&= \text{Tr}\Biggl(\Pi_x(\bm{\theta}) U_i\left(\theta_i+\frac{\pi}{2}\right)\rho(\bm{\theta})U^\dagger_i\left(\theta_i+\frac{\pi}{2}\right)\Biggr) - \text{Tr}\Biggl(\Pi_x(\bm{\theta}) U_i\left(\theta_i-\frac{\pi}{2}\right)\rho(\bm{\theta})U^\dagger_i\left(\theta_i-\frac{\pi}{2}\right)\Biggr) 
\nonumber
\\
&= q_{\bm{\theta}+\frac{\pi}{2}\bm{e}_i}(x)-q_{\bm{\theta}-\frac{\pi}{2}\bm{e}_i}(x).
\end{align}
Thus, the gradient formula can be compactly expressed as:
\begin{align}
&\frac{\partial \mathcal{L}(\bm{\theta})}{\partial \theta_i} = -\sum_x \frac{p(x)}{q_{\bm{\theta}}(x)}\left(q_{\bm{\theta}+\frac{\pi}{2}\bm{e}_i}(x) - q_{\bm{\theta}-\frac{\pi}{2}\bm{e}_i}(x)\right), \\
&\frac{\partial\mathcal{L}(\bm{\theta})}{\partial \theta_i} = \frac{\sum_x\sqrt{\frac{p(x)}{q_{\bm{\theta}}(x)}}\biggl(q_{\bm{\theta}+\frac{\pi}{2}\bm{e}_i}(x)-q_{\bm{\theta}-\frac{\pi}{2}\bm{e}_i}(x)\biggr)}{2\sqrt{1-\langle \sqrt{p},\sqrt{q_{\bm{\theta}}}\rangle^2}}.
\end{align}
This procedure simplifies the gradient evaluation to a series of tractable quantum measurements.

\clearpage
\section{Experiment results} 
\label{Appendix:experiment_results}
\begin{figure}[h]
    \centering
    \begin{subfigure}{0.3\textwidth}
        \centering
        \includegraphics[width=\textwidth]{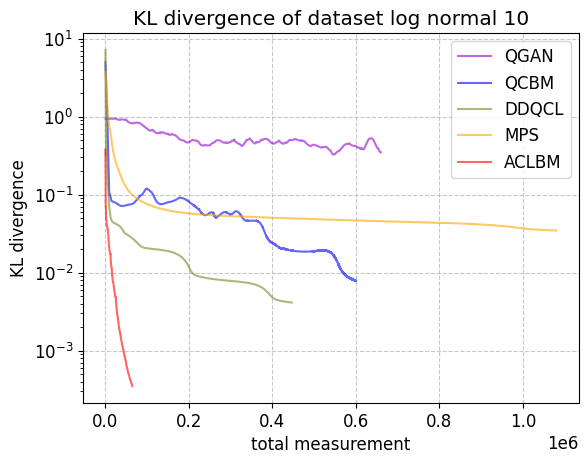}
    \end{subfigure}
    \begin{subfigure}{0.3\textwidth}
        \centering
        \includegraphics[width=\textwidth]{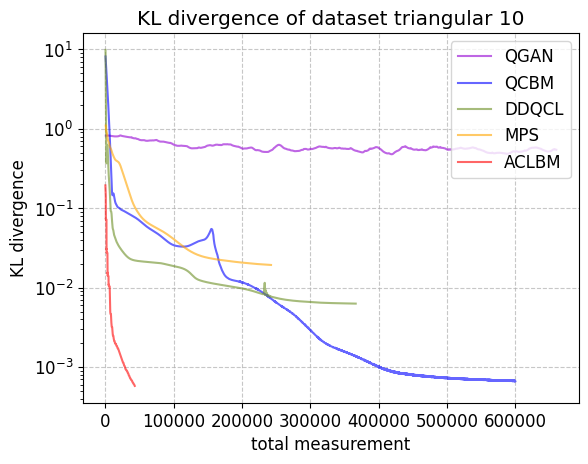}
    \end{subfigure}
    \begin{subfigure}{0.3\textwidth}
        \centering
        \includegraphics[width=\textwidth]{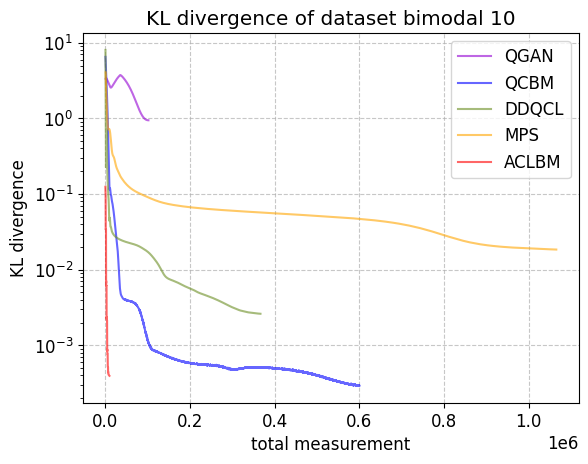}
    \end{subfigure}
    \begin{subfigure}{0.3\textwidth}
        \centering
        \includegraphics[width=\textwidth]{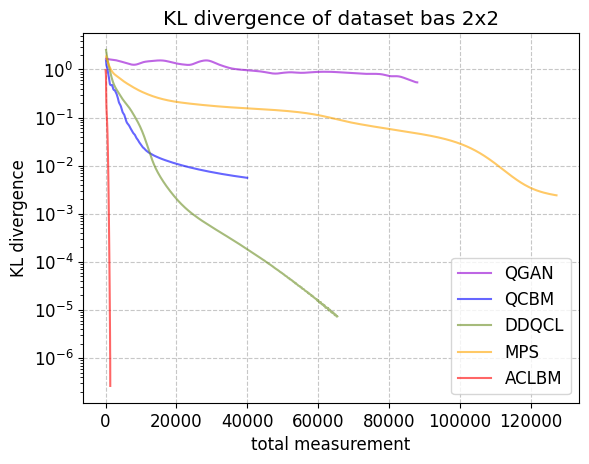}
    \end{subfigure}
    \begin{subfigure}{0.3\textwidth}
        \centering
        \includegraphics[width=\textwidth]{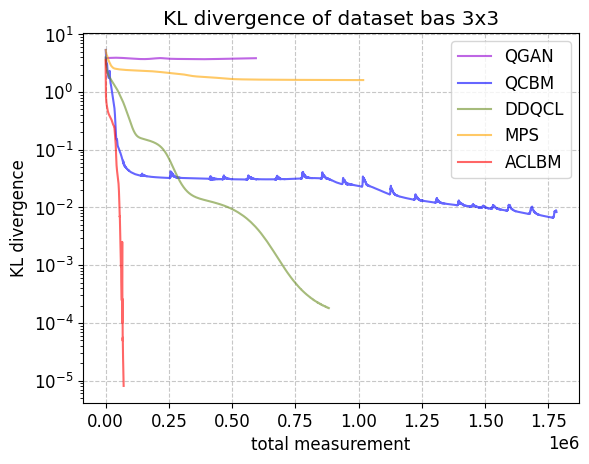}
    \end{subfigure}
    \begin{subfigure}{0.3\textwidth}
        \centering
        \includegraphics[width=\textwidth]{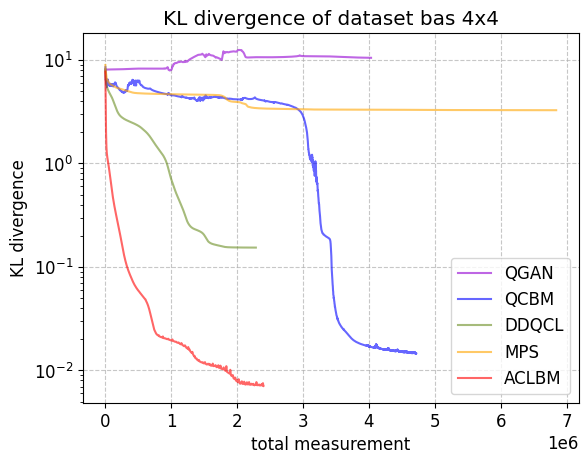}
    \end{subfigure}
    \begin{subfigure}{0.3\textwidth}
        \centering
        \includegraphics[width=\textwidth]{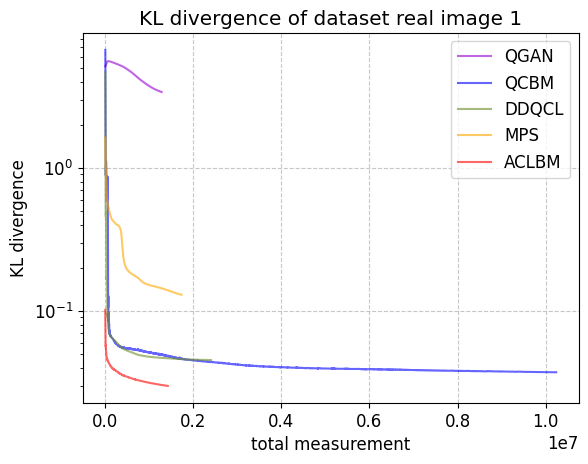}
    \end{subfigure}
    \begin{subfigure}{0.3\textwidth}
        \centering
        \includegraphics[width=\textwidth]{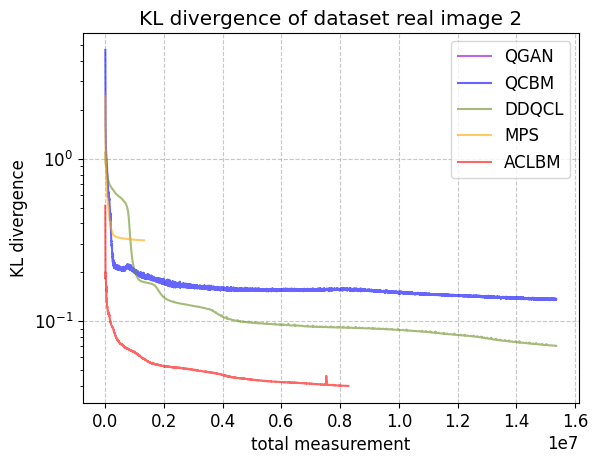}
    \end{subfigure}
    \begin{subfigure}{0.3\textwidth}
        \centering
        \includegraphics[width=\textwidth]{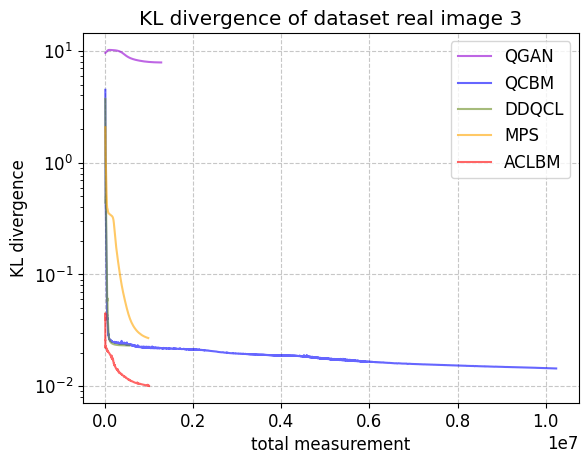}
    \end{subfigure}
    \caption{Benchmark Comparisons (KL divergence vs. total measurement) of different models on different datasets. Benchmark comparisons (KL divergence vs. total measurements) of different models across various datasets. The first row shows results for the log-normal 10, triangular 10, and bimodal 10 distributions. The second row displays results for the BAS $2\times 2$, BAS $3\times 3$, BAS $4\times 4$ datasets. The third row presents results for real images 1, 2 and 3. The number of measurements is calculated as follows: for QGAN, QCBM, DDQCL, and MPS, the number of parameters remains fixed, so the number of measurements in each epoch is 2 $\times$ (number of parameters), as two measurements are required to evaluate the gradient of a single parameter. For ACLBM, however, the parameters differ in each iteration, and we also account for the measurement cost in the operator selection phase. Thus, in each iteration, the measurements for operator selection are 2 $\times$ (number of operators in the pool), and in each epoch, the measurements are 2 $\times$ (current number of parameters).}
    \label{fig:KL_measurement_count}
\end{figure}
\clearpage

\input{experiment_results}
\clearpage 

\section{Quantum Mutual Information and Entanglement of Formation}
\label{Appendix:Quantum Mutual Information and Entanglement of Formation}
Quantum mutual information (QMI) is a measure of the total correlation between two subsystems of a quantum state. It quantifies not only classical correlations but also encompasses quantum entanglement, providing a complete picture of the correlations present in a quantum system. The QMI of a bipartite state \(\rho_{AB}\) is defined as:
\begin{equation}
I(A:B) = S(\rho_A) + S(\rho_B) - S(\rho_{AB}),
\end{equation}
where \(S(\rho)\) denotes the von Neumann entropy of the state \(\rho\), and \(\rho_A\) and \(\rho_B\) are the reduced density matrices of subsystems A and B, respectively.

Entanglement of formation (EOF)~\cite{wootters1998entanglement, wootters2001entanglement} is a conceptually distinct measure that specifically quantifies the resources needed to create a given entangled state. For a bipartite quantum state \(\rho_{AB}\), the EOF is the minimum average entanglement of an ensemble of pure states that represents \(\rho_{AB}\). Formally, if \(\rho_{AB} = \sum_i p_i |\psi_i\rangle \langle\psi_i|\), where \(|\psi_i\rangle\) are pure states and \(p_i\) are the probabilities of the states, the EOF is given by:
\begin{equation}
E_F(A:B) = \min \sum_i p_i E(|\psi_i\rangle),
\end{equation}
where \(E(|\psi_i\rangle)\) is the entanglement entropy of the state \(|\psi_i\rangle\).

For two-qubit systems, EOF has a closed-form expression, which is derived from the measure known as concurrence. The concurrence, \( C \), for a pair of qubits in a mixed state \( \rho \), is calculated as:
\begin{equation}
C(\rho) = \max\{0, \lambda_1 - \lambda_2 - \lambda_3 - \lambda_4\},
\end{equation}
where \( \lambda_i \) (for \( i=1,2,3,4 \)) are the square roots of the eigenvalues, in descending order, of the non-Hermitian matrix \( \rho \tilde{\rho} \). Here, \( \tilde{\rho} \) is the spin-flipped state of \( \rho \):
\begin{equation}
\tilde{\rho} = (\sigma_y \otimes \sigma_y) \rho^* (\sigma_y \otimes \sigma_y),
\end{equation}
and \( \rho^* \) denotes the complex conjugate of \( \rho \) in a fixed basis such as \( \{|00\rangle, |01\rangle, |10\rangle, |11\rangle\} \).

The EOF can then be expressed as a function of concurrence:
\begin{equation}
E_F(\rho) = h\left(\frac{1 + \sqrt{1 - C(\rho)^2}}{2}\right),
\end{equation}
where \( h(x) = -x \log_2 x - (1 - x) \log_2 (1 - x) \) represents the binary entropy function.

This relationship provides a direct method to calculate EOF from the density matrix of two-qubit states, offering a valuable tool for quantifying entanglement.

In the case of partitioned systems, where we focus on a two-qubit subsystem \(A, B\) within a larger system that includes \(C\), the bipartite state \(\rho_{AB}\) can be obtained by tracing out subsystem \(C\):
\begin{equation}
    \rho_{AB}=\text{Tr}_{C}(\rho_{ABC}).
\end{equation}

\section{Error Bound of the Approximated Time Evolution Operator}
\label{Appendix:Error_bound_time_evolution}
In this section, we derive the error bound discussed in Section~\ref{section:Hamiltonian_Simulation}. We revisit the problem: given the desired distribution $p_j$ for $j \in \{1, \ldots, d\}$ corresponding to the coefficients in the Linear Combination of Unitaries (LCU) decomposition $H = \sum_{j} \alpha_j V_j$, where $p_j \coloneqq \alpha_j/\alpha$ and $\alpha = \sum_{j=1}^{d} |\alpha_j|$, if we can generate a distribution $q_j$ that approximates the distribution $p_j$ up to an error $\delta$ in terms of KL divergence, i.e.,
\begin{equation}
    D_{\text{KL}}(p \Vert q) = \sum_{j=1}^d p_j \log\frac{p_j}{q_j} \leq \delta,
\end{equation}
then we can construct an approximated block encoding $H' = \sum q_j V_j$. To evaluate their discrepancy, we usually quantify the error in terms of the spectral norm of their difference:
\begin{align}
    \left\Vert H/\alpha - H'\right\Vert = \left\Vert \sum_{j=1}^{d} (p_j - q_j) V_j \right\Vert \leq \sum_{j=1}^d |p_j - q_j| \left\Vert V_j \right\Vert = \sum_{j=1}^d |p_j - q_j|,
\end{align}
where the inequality follows from the triangle inequality of spectral norms, and the last equality is due to the fact that the spectral norms of unitary operators are 1. Therefore, the spectral norm of the difference between the two operators $H/\alpha$ and $H'$ is actually upper bounded by twice the total variation distance $\delta(p, q) = \frac{1}{2}\sum_{j=1}^{d} |p_j - q_j|$. To relate the two quantities, $D_{\text{KL}}(p \Vert q)$ and $\delta(p, q)$, we further apply Pinsker's inequality:
\begin{equation}
    \delta(p, q) \leq \sqrt{\frac{1}{2} D_{\text{KL}}(p \Vert q)} \hspace{0.25cm} \Rightarrow \hspace{0.25cm} \left\Vert H/\alpha - H'\right\Vert \leq \sqrt{2\delta}.
\end{equation}

The next step is to quantify the error propagation as we implement the time evolution operator $e^{iHt}$. A well-studied result, shown in~\cite{chakraborty2018power}, provides insight into this, and we invoke their lemma here:
\begin{lemma}[\cite{chakraborty2018power} Lemma 50]
\label{eq:time_evolution_bound}
    Let $t\in\mathbb{R}$ and $H, H'\in\mathbb{C}^{n\times n}$ Hermitian operators, then
    \begin{equation}
        \left\Vert e^{itH}-e^{itH'}\right\Vert\leq\vert t \vert\Vert H - H'\Vert.
    \end{equation}
\end{lemma}
Using Eq.~(\ref{eq:time_evolution_bound}), we can derive the error scaling of the time evolution operator in terms of the error in KL-divergence:
\begin{equation}
    \left\Vert e^{iHt} - e^{iH'\alpha t} \right\Vert \leq \sqrt{2\delta}\alpha t
\end{equation}
Considering the case where the time evolution operator can only be approximately implemented by an algorithm, we can further quantify the error relative to the true time evolution. Suppose an algorithm implements an operator $U'$ that achieves an $\epsilon$-approximation to $e^{iH'\alpha t}$ by querying the approximated block-encoding $H'$, such that $\left\Vert e^{iH'\alpha t} - U'\right\Vert\leq \epsilon$.  Then, the error relative to the true time evolution operator $e^{iHt}$ is 
\begin{equation}
    \left\Vert e^{iHt}-U'\right\Vert\leq \left\Vert e^{iHt} - e^{iH'\alpha t}\right\Vert + \left\Vert e^{iH'\alpha t} - U'\right\Vert\leq \sqrt{2\delta}\alpha t + \epsilon
\end{equation}

\section{Error Bound of the Approximated Expectation Value}
\label{Appendix:Error Bound of the Approximated Expectation Value}
In this section, we quantify the error in the expectation value of the function $f(x)$. This is necessary because we can only access the approximated oracle $\mathcal{A}'$, which implements the state preparation as follows:
\begin{equation}
    \mathcal{A}'|0^n\rangle = \sum_{x=1}^{2^n-1} \sqrt{q(x)} |x\rangle.
\end{equation}
When applying this oracle within the Quantum Amplitude Estimation (QAE) algorithm, it is equivalent to estimating the expectation value of the function $f(x)$ with respect to the probability distribution $q(x)$. If the two probability distributions $p(x)$ and $q(x)$ are close, in terms of the KL-divergence $D_{\text{KL}}(p\Vert q)\leq \delta$, then the error in the expectation value can be bounded as:
\begin{align}
    \left\vert\sum_{x=0}^{2^n-1}p(x)f(x)-\sum_{x=0}^{2^n-1}q(x)f(x)\right\vert &\leq \sum_{x=0}^{2^n-1}\left\vert p(x)-q(x)\right\vert \cdot \vert f(x) \vert
    \nonumber\\
    &\leq \Vert p - q \Vert_2 \cdot \Vert f \Vert_2
    \nonumber\\
    &\leq \Vert p - q \Vert_1 \cdot \Vert f \Vert_2
    \nonumber\\ 
    &\leq \sqrt{2\delta} \cdot \Vert f \Vert_2,
\end{align}
where we utilize Pinsker's inequality in the final step. The notation $\Vert f\Vert_p$ is defined as $\left(\sum_{x=0}^{2^n-1}\vert f(x)\vert^p\right)^{1/p}$.

\end{widetext}
\end{document}

%% file: experiment_results.tex
\setlength{\arrayrulewidth}{0.3mm}
\setlength{\tabcolsep}{0.6cm}
\renewcommand{\arraystretch}{1.5}

\begin{table*}[ht]
    \centering
    \begin{tabular}{|c|c|c|c|c|c|}
        \hline
        \multicolumn{6}{|c|}{Experiment results (KL divergence)} \\
        \hline
        Datasets & QGAN & QCBM & DDQCL & MPS & ACLBM \\ 
        \hline
        log normal 3 & $8.25\times10^{-3}$ & $5.23\times10^{-4}$ & $1.14\times10^{-3}$ & $2.11\times10^{-6}$ & $2.55\times10^{-6}$  \\
        \hline
        bimodal 3 & $5.77\times10^{-3}$ & $2.23\times10^{-7}$ & $1.29\times10^{-5}$ & $5.45\times10^{-5}$ & $5.38\times10^{-6}$ \\
        \hline
        triangular 3 & $4.81\times10^{-2}$ & $5.54\times10^{-4}$ & $2.20\times10^{-5}$ & $7.61\times10^{-5}$ & $2.20\times10^{-6}$ \\
        \hline
        log normal 10 & $3.48\times10^{-1}$ & $7.83\times10^{-3}$ & $4.13\times10^{-3}$ & $3.46\times10^{-3}$ & $3.25\times10^{-4}$ \\
        \hline
        bimodal 10 & $5.93\times10^{-1}$ & $2.92\times10^{-4}$ & $2.61\times10^{-3}$ & $1.84\times10^{-2}$ & $3.95\times10^{-4}$ \\
        \hline
        triangular 10 & $5.45\times10^{-1}$ & $6.64\times10^{-4}$ & $6.30\times10^{-3}$ & $1.93\times10^{-2}$ & $5.80\times10^{-4}$ \\
        \hline
        bas 2$\times$2 & $5.34\times10^{-1}$ & $5.59\times10^{-3}$ & $7.36\times10^{-6}$ & $2.42\times10^{-3}$ & $2.63\times10^{-7}$ \\
        \hline
        bas 3$\times$3 & $3.86$ & $8.34\times10^{-3}$ & $1.81\times10^{-4}$ & $1.61$ & $8.09\times10^{-6}$ \\
        \hline
        bas 4$\times$4 & $10.46$ & $1.46\times10^{-2}$ & $1.54\times10^{-1}$ & $3.26$ & $7.09\times10^{-3}$ \\
        \hline
        real image 1 & $3.52$ & $3.72\times10^{-2}$ & $4.53\times10^{-2}$ & $1.30\times10^{-1}$ & $2.99\times10^{-2}$ \\
        \hline
        real image 2 & $2.96$ & $1.37\times10^{-1}$ & $7.06\times10^{-2}$ & $3.15\times10^{-1}$ & $4.00\times10^{-2}$ \\
        \hline
        real image 3 & $4.59$ & $1.76\times10^{-2}$ & $2.31\times10^{-2}$ & $2.70\times10^{-2}$ & $1.00\times10^{-2}$ \\
        \hline
    \end{tabular}
    \caption{Experimental results across various datasets and benchmark models.}
    \label{tab:exp_result(KL_divergence)}
\end{table*}

\begin{table*}[ht]
    \centering
    \begin{tabular}{|c|c|c|c|c|c|}
        \hline
        \multicolumn{6}{|c|}{Experiment results (loss function)} \\
        \hline
        Datasets & QGAN & QCBM & DDQCL & MPS & ACLBM \\ 
        \hline
        log normal 3 & $0.6941 / 0.6931$ & $3.03\times10^{-7}$ & $1.14\times10^{-3}$ & $2.11\times10^{-6}$ & $2.55\times10^{-6}$  \\
        \hline
        bimodal 3 & $0.6930 / 0.6931$ & $3.78\times10^{-10}$ & $1.29\times10^{-5}$ & $5.45\times10^{-5}$ & $5.38\times10^{-6}$ \\
        \hline
        triangular 3 & $0.7186/0.6851$ & $1.88\times10^{-7}$ & $2.20\times10^{-5}$ & $7.61\times10^{-5}$ & $2.20\times10^{-6}$ \\
        \hline
        log normal 10 & $0.7571/0.6762$ & $-17.72$ & $4.13\times10^{-3}$ & $3.46\times10^{-3}$ & $3.25\times10^{-4}$ \\
        \hline
        bimodal 10 & $0.6666/0.7263$ & $-19.84$ & $2.61\times10^{-3}$ & $1.84\times10^{-2}$ & $3.95\times10^{-4}$ \\
        \hline
        triangular 10 & $0.7702/0.6670$ & $-20.57$ & $6.30\times10^{-3}$ & $1.93\times10^{-2}$ & $5.80\times10^{-4}$ \\
        \hline
        bas 2$\times$2 & $0.7298/0.6932$ & $5.44\times10^{-6}$ & $7.36\times10^{-6}$ & $2.42\times10^{-3}$ & $2.63\times10^{-7}$ \\
        \hline
        bas 3$\times$3 & $0.6967/0.6905$ & $1.63\times10^{-5}$ & $1.81\times10^{-4}$ & $1.61$ & $8.09\times10^{-6}$ \\
        \hline
        bas 4$\times$4 & $0.0054/13.0612$ & $1.04\times10^{-5}$ & $1.54\times10^{-1}$ & $3.26$ & $7.09\times10^{-3}$ \\
        \hline
        real image 1 & $0.6943/0.6900$ & $-18.67$ & $4.53\times10^{-2}$ & $1.30\times10^{-1}$ & $2.99\times10^{-2}$ \\
        \hline
        real image 2 & $0.0003/13.7982$ & $-17.31$ & $7.06\times10^{-2}$ & $3.15\times10^{-1}$ & $4.00\times10^{-2}$ \\
        \hline
        real image 3 & $0.0300/13.3365$ & $-19.86$ & $2.31\times10^{-2}$ & $2.70\times10^{-2}$ & $1.00\times10^{-2}$ \\
        \hline
    \end{tabular}
    \caption{Experimental results across various datasets and benchmark models. Different loss functions are employed for different types of data distributions. For QCBM, the logarithmic MMD loss is used on the 10-qubit generic distribution dataset and the image dataset to enhance performance. For ACLBM, the Fisher-Rao metric is applied to the image dataset to capture intricate data distributions more effectively than the KL divergence.}
    \label{tab:exp_result(loss_function)}
\end{table*}

\begin{table*}[ht]
    \centering 
    \begin{tabular}{|c|c|c|c|c|}
    \hline
    \multicolumn{5}{|c|}{Resources of ACLBM} \footnote{The circuit depth is defined as the number of layers, where each layer contains gates that can be applied in parallel. For example, a circuit implementing the Pauli rotation $e^{-i\theta X \otimes Y / 2}$ involves 5 parallel layers, while $CR_Y(\theta)$ and $R_Y(\theta)$ each involve only a single parallel layer. Circuit information, including the number of one- and two-qubit gates and circuit depth, is calculated using PennyLane’s ``\href{https://docs.pennylane.ai/en/stable/introduction/inspecting_circuits.html}{qml.spec}” method.
    }\\
    \hline
    Datasets & Parameters & One qubit gates & Two qubit gates & Circuit depth \\
    \hline
    log normal 3 & 12 & 42 & 22 & 49 \\
    \hline
    bimodal 3 & 27 & 70 & 45 & 96 \\
    \hline 
    triangular 3 & 12 & 37 & 21 & 46 \\
    \hline 
    log normal 10 & 99 & 372 & 195 & 256 \\
    \hline
    log normal 10 -- 1 & 96 & 364 & 192 & 238 \\
    \hline 
    log normal 10 -- 2 & 105 & 397 & 208 & 247 \\
    \hline 
    log normal 10 -- 3 & 93 & 341 & 184 & 250 \\
    \hline
    bimodal 10 & 33 & 96 & 57 & 63 \\
    \hline 
    triangular 10 & 75 & 224 & 133 & 174 \\
    \hline
    bas 2$\times$2 & 10 & 30 & 20 & 38 \\
    \hline 
    bas 3$\times$3 & 80 & 132 & 124 & 110 \\
    \hline
    bas 4$\times$4 & 400 & 720 & 640 & 407 \\
    \hline
    bas 4$\times$4 $(N_o=4)$ & 64 & 186 & 126 & 161 \\
    \hline 
    bas 4$\times$4 $(N_o=40)$ & 200 & 465 & 351 & 331 \\
    \hline
    bas 4$\times$4 $(N_o=175, r=0.5)$ & 350 & 775 & 541 & 427 \\
    \hline
    real image 1 & 432 & 1517 & 861 & 755 \\
    \hline 
    real image 1 $(r=0.1)$ & 399 & 1180 & 653 & 586 \\
    \hline 
    real image 1 $(r=0.2)$ & 177 & 599 & 298 & 334 \\
    \hline 
    real image 1 $(r=0.3)$ & 102 & 310 & 170 & 256 \\
    \hline 
    real image 1 $(r=0.4)$ & 45 & 137 & 76 & 169 \\
    \hline 
    real image 1 $(r=0.5)$ & 57 & 165 & 96 & 213 \\
    \hline 
    real image 1 $(r=0.6)$ & 69 & 216 & 119 & 269 \\
    \hline 
    real image 1 $(r=0.7)$ & 9 & 18 & 13 & 25 \\
    \hline 
    real image 1 $(r=0.8)$ & 33 & 93 & 54 & 117 \\
    \hline 
    real image 1 $(r=0.9)$ & 15 & 34 & 23 & 47 \\
    \hline 
    real image 1 (remapped) & 102 & 411 & 191 & 312 \\
    \hline
    real image 2 & 750 & 2768 & 1498 & 1294 \\
    \hline
    real image 2 (remapped) & 33 & 125 & 66 & 99 \\
    \hline 
    real image 3 & 312 & 1031 & 619 & 545 \\
    \hline 
    real image 3 (remapped) & 63 & 225 & 120 & 187 \\
    \hline
    \end{tabular}
    \caption{Resources on each dataset of ACLBM}
\end{table*}